\documentclass{aa}  
\usepackage{natbib}
\bibpunct{(}{)}{;}{a}{}{,} 

\usepackage[english]{babel}           
\usepackage{graphicx,epsfig}
\usepackage{savesym}
\usepackage[fleqn]{amsmath}
\savesymbol{iint}
\savesymbol{iiint}
\savesymbol{iiiint}
\usepackage{txfonts}
\restoresymbol{TXF}{iint}
\restoresymbol{TXF}{iiint}
\restoresymbol{TXF}{iiiint}
\usepackage{lscape}
\usepackage{eucal}
\usepackage{pifont}
\usepackage{rotating}
\usepackage{multirow}
\usepackage{tikz}
\usepackage{longtable}

\DeclareMathOperator*{\argmax}{max}

\def \intens{\ifmmode{{\rm erg}\,\,{\rm cm}^{-2}\,\,{\rm s}^{-1}\,\,{\rm Hz}^{-1}\,\,{\rm sr}^{-1}}
                \else{${\rm erg}\,\,{\rm cm}^{-2}\,\,{\rm s}^{-1}\,\,{\rm Hz}^{-1}\,\,{\rm sr}^{-1}$}\fi}

\def \cc{\ifmmode{\,{\rm cm}^{-3}}\else{$\,{\rm cm}^{-3}$}\fi}
\def \ccubes{\ifmmode{\,{\rm cm}^{3}\,\,{\rm s}^{-1}}\else{$\,{\rm cm}^{3}\,\,{\rm s}^{-1}$}\fi}
\def \cq{\ifmmode{\,{\rm cm}^{-2}}\else{$\,{\rm cm}^{-2}$}\fi}
\def \mic{\ifmmode{\,\mu{\rm m}}\else{$\mu$m}\fi}
\def \eccs{\ifmmode{\,{\rm erg}\,{\rm cm}^{-3} {\rm s}^{-1}}\else{$\,{\rm
erg}\,{\rm cm}^{-3} {\rm s}^{-1}$}\fi}
\def \ecqs{\ifmmode{\,{\rm erg}\,{\rm cm}^{-2}\,{\rm s}^{-1}\,{\rm
sr}^{-1}}\else{$\,{\rm erg}\,{\rm cm}^{-2}\,{\rm s}^{-1}\,{\rm sr}^{-1}$}\fi}
\def \deg{\ifmmode{^{\circ}}\else{$^{\circ}$}\fi} 
\def \pc{\ifmmode{\,{\rm pc}}\else{$\,{\rm pc}$}\fi} 
\def \kms{\ifmmode{\,{\rm km}\,{\rm s}^{-1}}\else{km s$^{-1}$}\fi} 
\def \kmspc{\ifmmode{\,{\rm km}\,{\rm s}^{-1}\,{\rm pc}^{-1}}\else{km
s$^{-1}$ pc$^{-1}$}\fi} 
\def \MJysr{\ifmmode{\,{\rm MJy\,sr}^{-1}}\else{$\,{\rm MJy\,sr}^{-1}$}\fi} 
\def \Kkms{\ifmmode{\,{\rm K\,km\,s}^{-1}}\else{$\,{\rm K\,km\,s}^{-1}$}\fi} 
\def \twCO{\ifmmode{\rm ^{12}CO}\else{$\rm^{12}CO$}\fi} 
\def \thCO{\ifmmode{\rm ^{13}CO}\else{$\rm^{13}CO$}\fi} 
\def \twCN{\ifmmode{\rm ^{12}CN}\else{$\rm^{12}CN$}\fi} 
\def \thCN{\ifmmode{\rm ^{13}CN}\else{$\rm^{13}CN$}\fi} 
\def \HdCO{\ifmmode{\rm H_{2}CO}\else{$\rm H_{2}CO$}\fi} 
\def \twHdCO{\ifmmode{\rm H_{2}^{12}CO}\else{$\rm H_{2}^{12}CO$}\fi} 
\def \thHdCO{\ifmmode{\rm H_{2}^{13}CO}\else{$\rm H_{2}^{13}CO$}\fi} 
\def \twC{\ifmmode{\rm ^{12}C}\else{$\rm^{12}C$}\fi} 
\def \thC{\ifmmode{\rm ^{13}C}\else{$\rm^{13}C$}\fi} 
\def \Hp{\ifmmode{\rm H^+}\else{$\rm H^+$}\fi} 
\def \Cp{\ifmmode{\rm C^+}\else{$\rm C^+$}\fi} 
\def \Cpp{\ifmmode{\rm C^{++}}\else{$\rm C^{++}$}\fi} 
\def \Sp{\ifmmode{\rm C^+}\else{$\rm S^+$}\fi} 
\def \Spp{\ifmmode{\rm S^{++}}\else{$\rm S^{++}$}\fi} 
\def \CFp{\ifmmode{\rm CF^+}\else{$\rm CF^+$}\fi}
\def \CHp{\ifmmode{\rm CH^+}\else{$\rm CH^+$}\fi}
\def \CHdp{\ifmmode{\rm CH_2^+}\else{$\rm CH_2^+$}\fi}
\def \CHtp{\ifmmode{\rm CH_3^+}\else{$\rm CH_3^+$}\fi} 
\def \SHp{\ifmmode{\rm SH^+}\else{$\rm SH^+$}\fi}
\def \SH2p{\ifmmode{\rm SH_2^+}\else{$\rm SH_2^+$}\fi}
\def \twCHp{\ifmmode{\rm ^{12}CH^+}\else{$\rm^{12}CH^+$}\fi}
\def \thCHp{\ifmmode{\rm ^{13}CH^+}\else{$\rm^{13}CH^+$}\fi}
\def \CtH{\ifmmode{\rm C_2H}\else{$\rm C_2H$}\fi} 
\def \CthHt{\ifmmode{\rm C_3H_2}\else{$\rm C_3H_2$}\fi} 
\def \Htp{\ifmmode{\rm H_3^+}\else{$\rm H_3^+$}\fi} 
\def \HCOp{\ifmmode{\rm HCO^+}\else{$\rm HCO^+$}\fi} 
\def \HtOp{\ifmmode{\rm H_3O^+}\else{$\rm H_3O^+$}\fi} 
\def \HCfiN{\ifmmode{\rm HC_5N}\else{$\rm HC_5N$}\fi} 
\def \wat{\ifmmode{\rm H_2O}\else{$\rm H_2O$}\fi} 
\def \HdO{\ifmmode{\rm H_2O}\else{$\rm H_2O$}\fi} 
\def \OHp{\ifmmode{\rm OH^+}\else{$\rm OH^+$}\fi} 
\def \HdOp{\ifmmode{\rm H_2O^+}\else{$\rm H_2O^+$}\fi} 
\def \NHd{\ifmmode{\rm NH_2}\else{$\rm NH_2$}\fi} 
\def \NHtrois{\ifmmode{\rm NH_3}\else{$\rm NH_3$}\fi} 
\def \oxy{\ifmmode{\rm O_2}\else{$\rm O_2$}\fi} 
\def \HH{\ifmmode{\rm H_2}\else{$\rm H_2$}\fi}
\def \He{\ifmmode{\rm He}\else{$\rm He$}\fi}
\def \Jone{\ifmmode{\rm {(J=1--0)}}\else{{(J=1--0)}}\fi} 
\def \Jtwo{\ifmmode{\rm {(J=2--1)}}\else{{(J=2--1)}}\fi} 
\def \Jthr{\ifmmode{\rm {(J=3--2)}}\else{{(J=3--2)}}\fi} 
\def \Jfou{\ifmmode{\rm {(J=4--3)}}\else{{(J=4--3)}}\fi} 
\def \Jfiv{\ifmmode{\rm {J=4--3}}\else{{J=4--3}}\fi} 
\def \Ta{\ifmmode{\rm T_A}\else{$\rm T_A$}\fi} 
\def \Tas{\ifmmode{\rm T_A^*}\else{$\rm T_A^*$}\fi} 
\def \Tmb{\ifmmode{\rm T_{mb}}\else{$\rm T_{mb}$}\fi} 
\def \Tr{\ifmmode{\rm T_r}\else{$\rm T_r$}\fi} 
\def \Trs{\ifmmode{\rm T_r^*}\else{$\rm T_r^*$}\fi}
\def \TK{\ifmmode{T_{\rm K}}\else{$T_{\rm K}$}\fi} 
\def \TF{\ifmmode{T_{\rm F}}\else{$T_{\rm F}$}\fi} 
\def \TD{\ifmmode{T_{\rm D}}\else{$T_{\rm D}$}\fi} 
\def \NHt{\ifmmode{N_{\rm H}}\else{$N_{\rm H}$}\fi}
\def \NH{\ifmmode{N({\rm H})}\else{$N({\rm H})$}\fi}
\def \NH2{\ifmmode{N({\rm H}_2)}\else{$N({\rm H}_2)$}\fi}
\def \NCH{\ifmmode{N({\rm CH})}\else{$N({\rm CH})$}\fi}
\def \NHF{\ifmmode{N({\rm HF})}\else{$N({\rm HF})$}\fi}
\def \nH{\ifmmode{n_{\rm H}}\else{$n_{\rm H}$}\fi}
\def \nCO{\ifmmode{n({\rm CO})}\else{$n({\rm CO})$}\fi}
\def \nHF{\ifmmode{n({\rm HF})}\else{$n({\rm HF})$}\fi}
\def \nH2{\ifmmode{n({\rm H}_2)}\else{$n({\rm H}_2)$}\fi}
\def \nH{\ifmmode{n_{\rm H}}\else{$n_{\rm H}$}\fi}
\def \ngam{\langle n_{\gamma,\nu} \rangle}

\begin{document}

\title{A complete model of \CHp\ rotational excitation including radiative and chemical pumping
processes}

\author{
  B. Godard              \inst{1} \and
  J. Cernicharo          \inst{1}
}

\institute{
  Departamento de Astrof\'isica, Centro de Astrobiolog\'ia, CSIC-INTA, Torrej\'on de Ardoz, Madrid, Spain
  }

 \date{Received 2 August 2012 / Accepted 5 November 2012}

\abstract{} 
{Excitation of far-infrared and submillimetric molecular lines may originate from nonreactive 
collisions, chemical formation, or far infrared, near-infrared, and optical fluorescences. 
As a template, we investigate the impact of each of these processes on the excitation of the 
methylidyne cation \CHp\ and on the intensities of its rotational transitions recently detected 
in emission in dense photodissociation regions (PDRs) and in planetary nebulae.}
{We have developed a nonlocal thermodynamic equilibrium (non-LTE) excitation model that includes
the entire energy structure of \CHp, i.e. taking into account the pumping of its vibrational 
and bound and unbound electronic states by near-infrared and optical photons. The model 
includes the theoretical cross-sections of nonreactive collisions with H, \HH, He, and $e^-$, 
and a Boltzmann distribution is used to describe the probability of populating the excited levels 
of \CHp\ during its chemical formation by hydrogenation of \Cp. To confirm our results 
we also performed an extensive analytical study, which we use to predict the main excitation 
process of several diatomic molecules, namely HF, HCl, SiO, CS, and CO.}
{At densities $\nH=10^4$ \cc, the excitation of the rotational levels of \CHp\ is dominated by 
the radiative pumping of its electronic, vibrational, 
and rotational states if the intensities of the radiation field at $\sim 0.4$, $\sim 4$, and 
$\sim 300$ $\mu$m are stronger than $10^5$, $10^8$, and $10^4$ times those of the local 
interstellar radiation field (ISRF).
Below these values, the chemical pumping is the dominant source of excitation 
of the $J>1$ levels, even at high kinetic temperatures ($\sim 1000$ K). The far-infrared emission 
lines of \CHp\ observed in the Orion Bar and the NGC 7027 PDRs are consistent with the predictions 
of our excitation model assuming an incident far-ultraviolet (FUV) radiation field of $\sim 3 \times 
10^4$ (in Draine's unit) and densities of $\sim 5 \times 10^4$ and $\sim 2 \times 10^5$ \cc. 
In the case of NGC 7027, the estimate of the density is 10 to 100 times lower 
than those deduced by traditional excitation codes. Applying our model to other $X^1\Sigma^+$ 
ground state diatomic molecules, we find that HF, and SiO and HCl are the 
species the most sensitive to the radiative pumping of their vibrational and bound electronic 
states. In both cases, the minimal near-infrared and optical/UV radiation field 
intensities required to modify their rotational level populations are $\sim 10^3$ times 
those of the local ISRF at densities $\nH=10^4$ \cc. All these results point towards interstellar 
media with densities lower than previously established and cast doubts on 
the clumpiness of well-studied molecular clouds.}
{}


   \authorrunning{B. Godard et al.}
   \titlerunning{A complete model of \CHp\ rotational excitation including the radiative and chemical pumping processes}
   \maketitle
%

\section{Introduction}

Our knowledge of the physical conditions of interstellar and circumstellar molecular clouds has 
grown through the confrontation of the observed intensities and shapes of atomic and molecular lines 
with those predicted by radiative transfer models. The puzzle consists of simultaneously reproducing 
the emission profiles of various species observed in a single astronomical source with the minimal
number of physical components with different temperatures, densities, and velocity fields. However,
the reliability of the results strongly depends on the approximations of the radiative transfer 
models, and in particular, on the processes expected to drive the distribution of a molecule among 
its excited levels. 
For instance, \citet{Morris1975} and \citet{Carroll1981} have shown that considering only the
excitation by nonelastic collisions and the radiative transitions between the rotational levels 
of the ground vibrational state may lead to an overestimation (by several orders of magnitude) of 
the density of the main collisional partner or of the column density of the molecule.

Previous theoretical investigations of alternative excitation mechanisms have mainly focused on the 
capacity of near-infrared photons emitted by polycyclic hydrocarbon molecules (PAHs) to pump the 
molecules to an excited vibrational state in bright PDRs \citep{Scoville1980,Giard1997}. While this 
mechanism is negligible for certain transitions (e.g. HF ($1-0$) and HCl ($1-0$), \citealt{Agundez2011}), 
the near- and mid-infrared pumping of the vibrational states has been found to dominate the rotational 
excitation of several other species (e.g. \HdO, HNC, NH$_3$) in the circumstellar envelopes of 
AGB stars such as IRC10216 \citep{Gonzalez-Alfonso1999,Gonzalez-Alfonso2007,Agundez2008}. Similarly, 
the pumping of the electronic states by absorption of UV and optical photons followed by radiative 
decay has been proposed to play a major role in the rotational excitation of diatomic molecules 
\citep{Cernicharo1997}. For instance, this process has been studied to account for the 
near-infrared emission lines of \HH\ detected in the interstellar and circumstellar media (e.g. 
\citealt{Black1987,Habart2005}), the rovibrational emission lines of CO detected in molecular 
clouds and circumstellar disks (e.g. \citealt{Krotkov1980,Troutman2011}), the rotational
fine structure absorption lines of interstellar C$_2$ \citep{van-Dishoeck1982}, the 405 nm
band of interstellar C$_3$ observed in absorption towards HD210121 \citep{Roueff2002}, and
the rovibronic emission lines of \CHp\ detected in the blue spectrum of the Red Triangle 
nebula \citep{Balm1992}. Lastly, state-to-state chemistry, i.e. chemical reactions between 
the specific quantum states of the reagents and the products, may be involved in the excitation 
of a molecular species if the timescale of the reaction is comparable to that of nonreactive 
and nonelastic collisions \citep{Black1998}. Indeed, molecular excitation via the
chemical formation itself has already been proposed to populate the highest rotational levels
of interstellar C$_3$ \citep{Roueff2002}, and has recently been invoked by \citet{Goicoechea2011} 
as an alternative solution to explain the extended and intense rotational emission of OH 
detected in the Orion Bar PDR with the Herschel/PACS instrument.

Because all these processes could alter our concept of many astrophysical environments (from
interstellar and circumstellar PDRs to protoplanetary disks), it is essential to methodically 
explore their respective role on the rotational emission lines of each molecular species. While 
other diatomic molecules, such as OH and SiO, will be treated in forthcoming papers, we focus here 
on the methylidyne cation. Since the discovery of \CHp\ in the interstellar medium by 
\citet{Douglas1941}, the energy structure, spectroscopic properties, nonreactive 
collisional processes, and the chemistry of this molecular ion have been the subject of both 
experimental and theoretical studies (e.g. \citealt{Hakalla2006,Muller2010,Turpin2010,Gerlich1987,
Hierl1997} and reference therin). We have collated this information here into a 
concise paper with the aim to present a complete model of \CHp\ rotational excitation and propose 
an interpretation of its far-infrared emission lines recently observed with ISO and the Herschel 
Space Observatory.


\section{Pumping mechanisms of \CHp} \label{Pumping}

\begin{figure}[!hb]
\begin{center}
\includegraphics[width=9.0cm,angle=0]{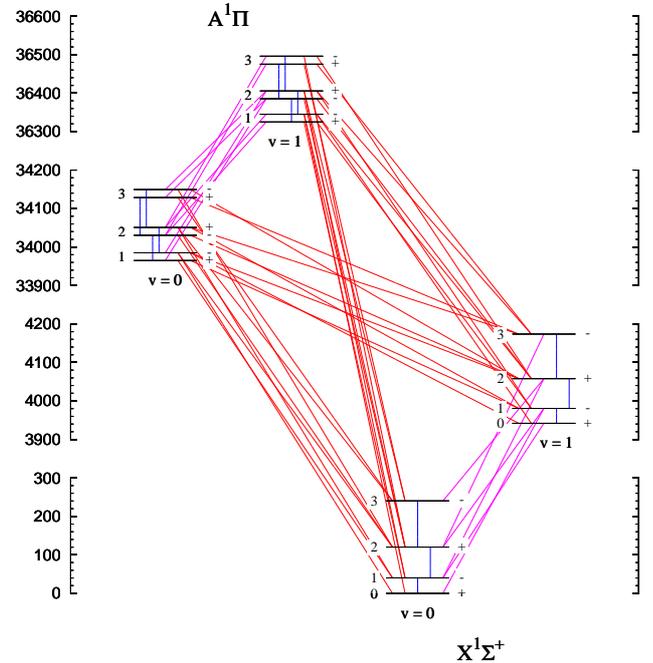}
\caption{Energies (in K) of the rovibrational levels of the first $X^1\Sigma^+$ and $A^1\Pi$ 
electronic states of \CHp. $\upsilon$-, $J$-, and $P$-values are given on the left, below, 
and on the right of each level. All allowed radiative transitions are displayed, including 
pure rotational (in blue), rovibrational (in purple), and electronic (in red) transitions.}
\label{Fig-Levels}
\end{center}
\end{figure}

\subsection{Energy structure}

\begin{table}[!!!ht]
\begin{center}
\caption{Equilibrium molecular constants (in cm$^{-1}$) of the two first electronic states of \CHp\ 
\citep{Hakalla2006}. Numbers in parenthesis are power of 10.}
\begin{tabular}{l r@{.}l l r@{.}l l}
\hline
constant        & \multicolumn{2}{c}{$X^1\Sigma^+$} &    & \multicolumn{2}{c}{$A^1\Pi$} &  \\
\hline
$T_e$           & \multicolumn{2}{c}{} &        &  2 & 41187262        & ($+4$) \\
$\omega_e$      & 2 & 8575609          & ($+3$) &  1 & 864402          & ($+3$) \\
$\omega_e x_e$  & 5 & 93179            & ($+1$) &  1 & 158317          & ($+2$) \\
$\omega_e y_e$  & 2 & 2534             & ($-1$) &  2 & 6301            &        \\
$B_e$           & 1 & 41774612         & ($+1$) &  1 & 1886774         & ($+1$) \\
$\alpha_e$      & 4 & 94739            & ($-1$) &  9 & 1629            & ($-1$) \\
$\gamma_e$      & 2 & 4904             & ($-3$) & -2 & 292             & ($-2$) \\
$\varepsilon_e$ & \multicolumn{2}{c}{} &        &  4 & 952             & ($-3$) \\
$D_e$           & 1 & 38914            & ($-3$) &  1 & 929606          & ($-3$) \\
$\beta_e$       &-2 & 66               & ($-5$) &  1 & 07331           & ($-4$) \\
$\delta_e$      & \multicolumn{2}{c}{} &        & -1 & 3123            & ($-5$) \\
$H_e$           & 1 & 2036             & ($-7$) & \multicolumn{2}{c}{} &        \\
$\alpha_{H_e}$  & 2 & 079              & ($-8$) & \multicolumn{2}{c}{} &        \\
$q_e$           & \multicolumn{2}{c}{} &        &  4 & 1018            & ($-2$) \\
$\alpha_{q_e}$  & \multicolumn{2}{c}{} &        & -3 & 135             & ($-3$) \\
$q_{D_e}$       & \multicolumn{2}{c}{} &        & -2 & 2               & ($-5$) \\
\hline
\end{tabular}
\label{TabEner}
\end{center}
\end{table}

Direct spectroscopic measurements of the pure rotational spectrum of \CHp\ are both chemically 
and technically difficult to achieve because (a) \CHp\ is an extremely reactive species and thus 
hard to maintain in detectable amounts, and (b) its rotational transitions lie at submillimeter 
and far-infrared wavelengths where the sensitivity of laboratory spectrometers has been limited 
until recently \citep{Pearson2006}. Hence, prior to \citet{Pearson2006}, who reported the first 
laboratory detection of the \CHp\ $J = 1-0$ line, all \CHp\ spectroscopic data were deduced from 
the measurements of the rovibronic bands $X^1\Sigma^+ - A^1\Pi$, $^1\Delta - ^1\Pi$, and 
$^3\Sigma - ^3\Pi$ (e.g. \citealt{Carre1968,Carrington1982,Hakalla2006}) and from the pure 
rotational spectrum observed towards NGC 7027 \citep{Cernicharo1997}.

Referring to Fig. \ref{Fig-Levels}, which displays the energy structure of the Douglas-Herzberg 
band system ($A^1\Pi$ - $X^1\Sigma^+$), the first $X^1\Sigma^+$ and $A^1\Pi$ electronic states 
of \CHp\ are split into a set of rovibrational levels. Their respective energies $E$ are described 
by the direct matrix elements of the effective Hamiltonian \citep{Brown1979}. Taking into account 
the spin-orbit coupling that results in splitting each rotational level in closed-spaced pairs of 
opposite parity ($e$ and $f$ levels, \citealt{Brown1975}), the energies of all levels can be 
computed as
\begin{equation}
\begin{array}{l}
E(\Lambda,v,J) = \\[5pt]
T_e(\Lambda) + \omega_e(\Lambda) (\upsilon+\frac{1}{2}) - \omega_e x_e(\Lambda) (\upsilon+\frac{1}{2})^2 + \omega_e y_e (\Lambda) (\upsilon+\frac{1}{2})^3 + \\[5pt]
\left[ B_e(\Lambda) - \alpha_e(\Lambda) (\upsilon+\frac{1}{2}) + \gamma_e(\Lambda) (\upsilon+\frac{1}{2})^2 + \varepsilon_e(\Lambda) (\upsilon+\frac{1}{2})^3 \right] \times \\[5pt]
\left( J (J+1) - \Lambda^2 \right) - \\[5pt]
\left[ D_e(\Lambda) + \beta_e(\Lambda) (\upsilon+\frac{1}{2}) + \delta_e(\Lambda) (\upsilon+\frac{1}{2})^2 \right] \left( J (J+1) - \Lambda^2 \right)^2 + \\[5pt]
\left[ H_e(\Lambda) - \alpha_{H_e}(\Lambda) (\upsilon+\frac{1}{2}) \right] \left( J (J+1) - \Lambda^2 \right)^3 \pm \\[5pt]
\frac{1}{2} \left[ q_e(\Lambda) + \alpha_{q_e}(\Lambda) (\upsilon+\frac{1}{2}) \right] \left( J (J+1) \right) \pm \\[5pt]
\frac{1}{2} q_{D_e}(\Lambda) \left( J (J+1) \right)^2,
\end{array}
\end{equation}
where $T_e(\Lambda)$, $\omega_e(\Lambda)$, etc. are the equilibrium molecular constants of 
the electronic state $\Lambda$, and where $+$ and $-$ refer to the $e$ and $f$ sublevels. 
Each level is therefore defined by its total orbital angular momentum $\Lambda$, vibrational 
momentum $\upsilon$, total angular momentum $J$, and parity $P$. With the equilibrium 
molecular constants derived by \citet{Hakalla2006} and given in Table \ref{TabEner}, the 
electronic, pure vibrational, and pure rotational level structures of \CHp\ are significantly 
separated in energy from one another as long as $\upsilon < 10$ and $J<13$.

\subsection{Nondissociative radiative processes}

\begin{table*}[!!ht]
\begin{center}
\caption{Einstein spontaneous emission coefficients $A_{\Lambda' \upsilon' J' \Lambda'' 
\upsilon'' J''}$ (in s$^{-1}$) computed for the $J',J''=1,0$ transitions. The data are taken
from \citet{Hakalla2006} for the \CHp\ $A^1\Pi - X^1\Sigma^+$ band system (top) and from 
\citet{Cheng2007} and \citet{Follmeg1987} for the \CHp\ $X^1\Sigma^+ - X^1\Sigma^+$ band 
system (bottom). Numbers in parenthesis are power of 10.}
\begin{tabular}{l @{\hspace{0.7cm}} l @{\hspace{0.2cm}} l @{\hspace{0.7cm}} l @{\hspace{0.2cm}} l 
                  @{\hspace{0.7cm}} l @{\hspace{0.2cm}} l @{\hspace{0.7cm}} l @{\hspace{0.2cm}} l
                  @{\hspace{0.7cm}} l @{\hspace{0.2cm}} l}
\multicolumn{11}{c}{$A^1\Pi - X^1\Sigma^+$ system} \\
\hline
$\upsilon ''$ \textbackslash $\upsilon '$ & \multicolumn{2}{c}{0} & \multicolumn{2}{c}{1} & \multicolumn{2}{c}{2} & \multicolumn{2}{c}{3} & \multicolumn{2}{c}{4} \\
\hline
0 & 4.2477 & ($+05$) & 2.9239 & ($+05$) & 1.4146 & ($+05$) & 6.2836 & ($+04$) & 2.8210 & ($+04$) \\
1 & 7.4827 & ($+04$) & 6.3350 & ($+04$) & 1.9902 & ($+05$) & 1.9663 & ($+05$) & 1.4058 & ($+05$) \\
2 & 8.6810 & ($+03$) & 6.5169 & ($+04$) & 1.5824 & ($+03$) & 3.7705 & ($+04$) & 9.7514 & ($+04$) \\
3 & 6.7337 & ($+02$) & 1.8585 & ($+04$) & 1.9247 & ($+04$) & 2.4184 & ($+04$) & 1.3421 & ($+03$) \\
4 & 4.1285 & ($+01$) & 2.7788 & ($+03$) & 1.7104 & ($+04$) & 1.5654 & ($+01$) & 1.2255 & ($+04$) \\
\hline
\\[0.4cm]
\multicolumn{11}{c}{$X^1\Sigma^+ - X^1\Sigma^+$ system} \\
\hline
$\upsilon ''$ \textbackslash $\upsilon '$ & \multicolumn{2}{c}{0} & \multicolumn{2}{c}{1} & \multicolumn{2}{c}{2} & \multicolumn{2}{c}{3} & \multicolumn{2}{c}{4} \\
\hline
0 & 6.4002 & ($-03$) & 5.6751 & ($-01$) & 4.6023 & ($-04$) & 1.3972 & ($-07$)$^{a}$ & 3.0814 & ($-11$)$^{a}$ \\
1 &        &         & 5.7869 & ($-03$) & 1.1568 & ($+00$) & 5.4871 & ($-02$)       & 1.1272 & ($-04$)$^{a}$ \\
2 &        &         &        &         & 5.2477 & ($-03$) & 1.8541 & ($+00$)       & 6.7870 & ($-01$) \\
3 &        &         &        &         &        &         & 4.7379 & ($-03$)       & 3.0926 & ($+00$) \\
4 &        &         &        &         &        &         &        &               & 4.2613 & ($-03$) \\
\hline
\end{tabular}
\begin{list}{}{}
\item[$^{a}$]  Extrapolated values assuming $M_{0 \upsilon' 0 0} = 0.65 \times {\rm exp}[-4.61\,\, \upsilon']$ 
$ea_0$, and $M_{0 \upsilon' 0 1} = 0.50 \times {\rm exp}[-3.38\,\, (\upsilon'-1) ]$ $ea_0$.
\end{list}
\label{TabSpec}
\end{center}
\end{table*}

In astrophysical environments, the level populations of a molecule are controlled by the spontenous 
and induced radiative processes and the collisional excitation. As shown in Fig. \ref{Fig-Levels}, 
we consider here all possible radiative transitions of  \CHp\ that meet the selection rules 
$\Delta \Lambda = 0, \,\, \pm 1$, $\Delta S = 0$, $\Delta J = 0 \,\, ({\rm if} \,\, 
\Delta \Lambda \ne 0), \,\, \pm 1$, and $P'P'' = -1$. The pure rotational transitions lie in the 
mid- and far-infrared part of the electromagnetic spectrum from 10 to 1000 $\mu$m, the rovibrational 
lines occupy the near- and mid-infrared domain between 1 and 10 $\mu$m, and the rovibronic 
transitions involve ultraviolet and optical photons between 0.3 and 0.8 $\mu$m.

With the conventions adopted by \citet{Larsson1983}, the absorption line oscillator strength $f$ and 
the Einstein spontaneous emission coefficient $A$ of a transition $\Lambda',\upsilon', J' \rightarrow 
\Lambda'',\upsilon'',J''$ of a diatomic molecule are given by 
\begin{equation}
f_{\Lambda' \upsilon' J' \Lambda'' \upsilon'' J''} = 
\frac{8\pi^2 m_e \nu}{3h e^2} M_{\Lambda' \upsilon' \Lambda'' \upsilon''}^2
\frac{S^{\Lambda' J'}_{\Lambda''J''}}{2J''+1},
\end{equation}
and
\begin{equation}
A_{\Lambda' \upsilon' J' \Lambda'' \upsilon'' J''} = 
\frac{8\pi^2 \nu^2 e^2}{m_e c^3} \frac{2J''+1}{2J'+1}
f_{\Lambda' \upsilon' J' \Lambda'' \upsilon'' J''},
\end{equation}
where $m_e$ and $e$ are the mass and the charge of the electron, $h$ is the Planck constant, 
$c$ is the speed of light, and $\nu$, $M$, and $S$ are the frequency, the transition 
dipole moment, and the H\"onl-London factors \citep{Herzberg1950}. Under the 
Born-Oppenheimer and the $r$-centroid approximations, the electronic and nuclear motions are 
separated and $M$ can be expressed as
\begin{equation}
M_{\Lambda' \upsilon' \Lambda'' \upsilon''}^2 = q_{\Lambda' \upsilon' \Lambda'' \upsilon''} 
R_e^2(\overline{r}_{\upsilon' \upsilon''}),
\end{equation}
where $\overline{r}$ is the $r$-centroid of the $\upsilon'- \upsilon''$ vibrational band, $q$ is
the Franck-Condon factor, and $R_e$ is the electronic transition moment.

Collecting the vibrational band oscillator strengths computed by \citet{Elander1977} and measured
by \citet{Weselak2009a}, and using the Franck-Condon factors and $r$-centroids obtained by 
\citet{Hakalla2006}, we found that the electronic transition moment of the $A^1\Pi - X^1\Sigma^+$ 
system has a clearly defined functional form (with a correlation coefficient of 0.97)
\begin{equation} \label{eqRe}
R_e(\overline{r}_{\upsilon' \upsilon''}) = 2.01 \times {\rm exp}(-1.66 \,\, \overline{r}_{\upsilon' \upsilon''}) \quad ea_0.
\end{equation}
Conversely, since we were unable to extract a simple relation between the transition dipole moment 
and the internuclear separation \citep{Chackerian1982} for the $X^1\Sigma^+ - X^1\Sigma^+$ system, 
we extrapolated $M_{0 \upsilon' 0 \upsilon''}$ from the values computed at lower $\upsilon'$ by 
\citet{Cheng2007} and \citet{Follmeg1987} assuming an exponential dependence on $\upsilon' - 
\upsilon''$. Table \ref{TabSpec} contains the corresponding Einstein spontaneous emission 
coefficients computed for the $J',J'' = 1,0$ transitions of the first vibrational bands of the
$A^1\Pi - X^1\Sigma^+$ and $X^1\Sigma^+ - X^1\Sigma^+$ systems.

\subsection{Nonreactive nonelastic collisional processes}

Since we aim to study the \CHp\ excitation mechanisms over a wide domain of parameters (i.e. 
describing different astronomical sources), and, for example, at kinetic temperature $T_{\rm K}$
close to the equivalent temperatures of the rovibrational transitions ($> 1000$ K), we include all 
nonreactive collisional excitation and de-excitation processes, even between two different 
rovibronic levels, and considering H, \HH, He, and e$^-$ as collision partners. Unfortunately, 
there are no experimental data available for the \CHp\ collisional rates, and the theoretical data 
are usually limited to the pure rotational levels.

For collisions with electrons, we adopted the rate functions of \TK\ obtained by \citet{Lim1999} 
for $\upsilon=0$ and $J$ up to 7. For collisions with neutrals, we used the \CHp-He collisional 
rates computed by \citet{Hammami2008,Hammami2009} and \citet{Turpin2010} for $\upsilon=0$ and 
$J$ up to 10, and then applied the rigid rotator approximation\footnote{Within the rigid rotator 
approximation, the collisional rates are proportional to $\mu^{-1/2}$, where $\mu$ is the 
relative mass of the system collider-\CHp.} to take into account collisions with H and \HH. 
For $300 {\rm K} \leqslant \TK \leqslant 2000 {\rm K}$, we used the data of \citet{Hammami2009}. 
Below this temperature range, we preferentially adopted the values computed by \citet{Turpin2010} 
rather than those of \citet{Hammami2008} because the former authors have found that including 
of the long-range charge-induced dipole potential of \CHp\ has an impact on the collisional 
rates at low temperature. Finally, for \TK\ higher than 2000 K, we extrapolated the de-excitation 
rates by collisions with neutrals from the values computed at lower temperature assuming a power 
law dependence on \TK.

These nonreactive collisional rates were used for all transitions with $\Lambda' = \Lambda''$ 
and $\upsilon' = \upsilon''$. In turn, the $\upsilon'=\upsilon'' + 1$ collisional rates were 
deduced from the theoretical cross-sections for inelastic collisions between \CHp\ and He
computed by \citet{Stoecklin2008} for $\upsilon',\upsilon''=1,0$ and $J',J''$ up to $0,5$.
For all other transitions occuring between two different vibrational or electronic levels, 
the downward rates were inferred from those of the pure rotational transitions assuming a 
powerlaw dependence on $(E'-E'')$, 
\begin{equation}
k^C_{\Lambda'\upsilon'J'\Lambda''\upsilon''J''} = (2J'+1)(2J''+1) \times a \left(\frac{E'-E''}{\TK} \right)^b,
\end{equation}
where $E'$ and $E''$ are the energies of the upper and lower levels and $a$ and $b$ are the 
best-fit coefficients of all available data at the temperature \TK. As a justification for 
this choice, the obtained scaling relations are found to describe the pure rotational 
transition rates within a factor of 3 and with a correlation coefficient ranging from 0.8 
to 0.9. We note, however, that they systematically underestimate the vibrational transition
rates estimated from the cross sections of \citet{Stoecklin2008} by a factor varying from 
70 to 5 for $10 \leqslant \TK \leqslant 70$ K, because the vibrational and rotational 
quenching cross sections have the same order of magnitude in the low collision energy limit 
\citep{Stoecklin2008}, and by a factor $< 5$ for $\TK > 70$ K. Nonetheless, since the 
discrepancies diminish when $\TK$ increases and because our subsequent analysis and 
predictions will mainly come from the high kinetic temperature domain, we consider our
fitting approach as valid.

\subsection{Reactive collisional processes and photodissociation}

Finally, since the chemical timescales of \CHp\ are comparable with those of the nonreactive
nonelastic collisions \citep{Bruderer2010}, we took into account the excitation and de-excitation 
of \CHp\ during its chemical formation and destruction. 

During the past decades, the advances of the cross molecular beam experiments and of the theoretical 
studies of chemical reaction dynamics (see the reviews of the field by \citealt{Casavecchia2000,
Levine2005}) have led to measurements and calculations of state-specific reaction rates 
for several neutral-neutral reactions (e.g. O($^3$P) + \HH($\upsilon=1$) $\rightarrow$ 
OH($\upsilon=1$) + H by \citealt{Balakrishnan2004}). For a few reactions these studies 
have even provided detailed diagrams of the most probable state of the products depending on the 
reagent excitation state and the collision energy (\citealt{Casavecchia2000} and references 
therein). 
Similarly, the recent improvements in the flowing afterglow apparatus and the ion trapping
techniques \citep{Gerlich2009} have led to the first experimental deductions of state-specific 
rate coefficients of $\Cp + \HH \rightleftarrows \CHp + {\rm H}$: depending on
the internal energy of \HH\ (up to $\upsilon=1$) for the forward reaction \citep{Hierl1997} 
and on the internal energy of \CHp\ (up to $J=3$) for the reverse process \citep{Plasil2011}.
In particular, \citet{Plasil2011} showed that the destruction rate of $\CHp(J=0)$ by reactive 
collisions with H is 2 to 30 times lower than that of rotationally excited \CHp.

To include the \CHp\ state-to-state chemistry, we followed the approach described by 
\citet{Black1998} and \citet{van-der-Tak2007}.
\begin{itemize}
\item[$\bullet$] We considered that $\CHp(\Lambda,\upsilon,J)$ is mainly destroyed by 
reactions with H, \HH, and $e^-$ or by photodissociation. Because of lack of information,
we also assumed that the three last reactions are independent of $\Lambda$, $\upsilon$, 
and $J$ by adopting generic reaction rates for all levels: $k^D_{\HH} = 1.2 \times 10^{-9}$ 
\ccubes, $k^D_{e^-} = 1.5 \times 10^{-7} (\TD/300\,{\rm K})^{-0.42}$ \ccubes, and 
$k^D_{\gamma} =\int \frac{4\pi}{h\nu} \sigma^\gamma(\nu) I_\nu d\nu$ s${^-1}$ for reactions 
with \HH\ and $e^-$ and for the photodissociation respectively \citep{McEwan1999,Mitchell1990,
van-Dishoeck2006}, where $\TD$ is the temperature of the reaction and $\sigma^\gamma(\nu)$ 
are the \CHp\ photodissociation cross-sections computed by \citet{Kirby1980}. Conversely, we 
adopted the state-specific rates of \citet{Plasil2011} for the destruction of \CHp\ by collisions 
with H atoms.
\item[$\bullet$] During the chemical formation of \CHp\, we assumed that the rotational level 
populations follow a Boltzmann distribution at the temperature of the reaction \TF:
\begin{equation}
k^F_{\Lambda,\upsilon,J} = k^F \,\, (2J+1) \times {\rm exp}(-E_{\Lambda,\upsilon,J}/\TF)\,/\,\mathcal{Q}(\TF),
\end{equation}
where $k^F$ is the state-averaged formation rate and $\mathcal{Q}(\TF)$ is the partition function 
of \CHp. 
\end{itemize}
The formation and destruction temperatures correspond to the collision energies of the reagents and
thus depend on their individual dynamics and on the kinetic temperature of the gas. However, since
our subsequent analysis is based on the results of static models at chemical equilibrium, we 
assumed in the following $\TD=\TF=\TK$, unless specified otherwise. Because the dynamical properties
of the gas are neglected, this hypothesis is a lower limit on the actual values of $\TD$ and $\TF$.


\subsection{Critical parameter analysis} \label{Section-Exploration}

In practice, all the above processes are not necessarily significant over the entire domain of 
parameters, namely the density of the gas $\nH$ (in \cc), the kinetic temperature \TK\ (in K), 
and the intensity of the radiation field at all wavelengths $I_\nu$ (in \intens). Moreover, if 
they are all included in the radiative transfer models, they can lead to computational errors 
when the populations of the high-energy levels are too small. It is therefore useful to make a 
preliminary exploration of the parameter domain and reduce, if necessary, the number of levels 
and excitation processes considered depending on the physical conditions of the gas. To do so, 
we used the analytical treatment of the main de-excitation mechanisms that we present in Appendix 
\ref{Appen-Critic}. In addition to the critical density $n_{{\rm H \,\,cri},i}$, we derived for 
each level $i$ of a molecule four dimensionless critical parameters: $X_{{\rm cri},i}$, 
$Y_{{\rm cri},i}$, $Z_{{\rm cri},i}$, and $S_{{\rm cri},i}$ which depend on the reactive 
and nonreactive collisional rates, the abundances relative to H of the collision partners, 
the Einstein spontaneous emission coefficients, and the intensities of the radiation field at 
the corresponding wavelengths.

Given the energies and the lifetimes of the $\Lambda \geqslant 1$ or $\upsilon \geqslant 1$ levels,
collisional de-excitation of vibrational and electronic levels are found to be always negligible 
compared to radiative de-excitation. More importantly, our approach shows that the parameter space 
is divided for each rotational level into six regions where the de-excitation occurs mainly by (1) 
reactive collisions, (2) nonreactive collisions, (3) pure rotational, (4) rovibrational, 
(5) rovibronic radiative transitions, and (6) photodissociation. For example, if 
$n(\HH)/\nH=[\HH]=0.5$ and $n(e^-)/\nH=[e^-]=10^{-4}$, the limits of the first five regions 
for \CHp\ in the level $\Lambda,\upsilon,J=0,0,1$ are $\TK \sim 800$ K, $\nH \sim 10^{7}$ \cc, 
$I_\nu(4\,\,\mu{\rm m}) \sim 2 \times 10^{-8}$ \intens, and $I_\nu(0.4\,\,\mu{\rm m}) \sim 7 
\times 10^{-12}$ \intens.

Since the excitation processes are not included, the critical parameter approach is insufficient 
to calculate the distribution of a molecule among its levels. However, it proves useful to point 
out the physical conditions for which several excitation and de-excitation mechanisms have to be 
taken into account and the level populations must be computed with non-LTE radiative transfer 
models.

\section{Radiative transfer modelling} \label{Madex}

\subsection{Method: the MADEX code}

\begin{table}[!h]
\begin{center}
\caption{Reference parameters used in Eqs. \ref{Eq-ISRF1} - \ref{Eq-ISRF7} to compute the 
interstellar radiation field at high Galactic latitude. Numbers in parenthesis are power of 10.}
\begin{tabular}{l l r r r r}
\hline
frequency domain & \multicolumn{1}{c}{index} & \multicolumn{1}{c}{$T$ (K)} & \multicolumn{1}{c}{$\tau$} & \multicolumn{1}{c}{$\nu$ (GHz)} & \multicolumn{1}{c}{$\beta$} \\
\hline
microwave        & CMB    &  2.7 &           &         &     \\
far-infrared     & fir    &   18 & 1.7 (-05) & 1.1 (3) & 2.0 \\
mid-infrared 1   & mi$_1$ &   50 & 2.5 (-08) & 2.9 (3) & 1.0 \\
mid-infrared 2   & mi$_2$ &  260 & 3.0 (-10) & 1.5 (4) & 1.5 \\
near-infrared    & nir    & 3000 & 6.0 (-13) & 1.8 (5) & 1.0 \\
\hline
\end{tabular}
\label{Tab-ISRF}
\end{center}
\end{table}

\begin{figure}[!h]
\begin{center}
\includegraphics[width=9.0cm,angle=0]{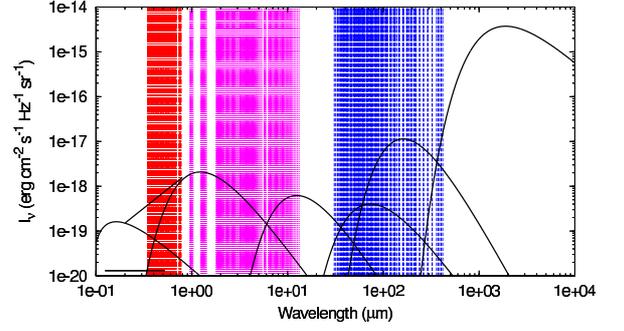}
\caption{Interstellar radiation field intensity $I_\nu$ at high Galactic latitude from Table 
\ref{Tab-ISRF} and Eqs. \ref{Eq-ISRF1} - \ref{Eq-ISRF7} as a function of the wavelength. The 
wavelengths of the 1725 allowed pure rotational, rovibrational, and rovibronic radiative 
transitions of \CHp\ are also displayed in blue, purple, and red, taking into account the 
first two, five, and thirteen electronic, vibrational, and rotational levels of \CHp.}
\label{Fig-ISRF}
\end{center}
\end{figure}

To study the individual impact of all pumping processes and compute the steady-state
level populations of \CHp\ in several astrophysical environments, we improved and updated the
MADEX (MADrid molecular spectroscopy EXcitation) excitation code \citep{Cernicharo2012}. MADEX is 
a radiative transfer model based on the multi-shell high velocity gradients (LVG) formalism 
\citep{Goldreich1974}. The physical 
model consists of a spherical (or plan parallel) cloud of gas and dust defined by the radial 
(or perpendicular) profiles of the density $\nH$; the \HH\, He, and $e^-$ abundances relative 
to H, $[{\rm He}]$, $[\HH]$, and $[e^-]$; the \CHp\ column density $N(\CHp)$; the kinetic 
temperature $\TK$; the expansion velocity of the cloud along the line of sight $\upsilon_{\rm 
exp}$; and the dust abundance and temperature $n_d$ and $T_d$.

Gas and dust are pervaded by an internal radiation field emitted by a central star with a radius 
$R_{\star}$ and an effective temperature $T_{\star}$, and an external radiation field defined by
its specific intensity $I_\nu$. To cover a broad range of the electromagnetic spectrum, the external 
radiation field is modelled in a simple and versatile manner as
\begin{align}
\label{Eq-ISRF1}
I_\nu & = \frac{2 h \nu^{3}}{c^2} \,\, \frac{1}{{\rm exp}(h \nu / k T_{\rm CMB}) - 1} \\
\label{Eq-ISRF2}
      & + \frac{2 h \nu^{3}}{c^2} \,\, \frac{\chi_{\rm fir }}{{\rm exp}(h \nu / k T_{\rm fir }) - 1} \,\, \left[ 1 - {\rm exp}(-\tau_{\rm fir }(\nu/\nu_{\rm fir })^{\beta_{\rm fir }}) \right]  \\
\label{Eq-ISRF3}
      & + \frac{2 h \nu^{3}}{c^2} \,\, \frac{\chi_{\rm mi_1}}{{\rm exp}(h \nu / k T_{\rm mi_1}) - 1} \,\, \left[ 1 - {\rm exp}(-\tau_{\rm mi_1}(\nu/\nu_{\rm mi_1})^{\beta_{\rm mi_1}}) \right]  \\
\label{Eq-ISRF4}
      & + \frac{2 h \nu^{3}}{c^2} \,\, \frac{\chi_{\rm mi_2}}{{\rm exp}(h \nu / k T_{\rm mi_2}) - 1} \,\, \left[ 1 - {\rm exp}(-\tau_{\rm mi_2}(\nu/\nu_{\rm mi_2})^{\beta_{\rm mi_2}}) \right]  \\
\label{Eq-ISRF5}
      & + \frac{2 h \nu^{3}}{c^2} \,\, \frac{\chi_{\rm nir }}{{\rm exp}(h \nu / k T_{\rm nir }) - 1} \,\, \left[ 1 - {\rm exp}(-\tau_{\rm nir }(\nu/\nu_{\rm nir })^{\beta_{\rm nir }}) \right]  \\
\label{Eq-ISRF6}
      & + \chi_{\rm opt} \,\,9.4 \,\, 10^{6}\,\,\nu^{-1.7} \,\, \text{if $3.7 \times 10^{14} \leqslant \nu \leqslant 1.5 \times 10^{15}$ GHz} \\
\label{Eq-ISRF7}
      & + \frac{\chi_{\rm uv}}{4 \pi} \left[ \frac{6.36\,\,10^{-17} \nu^2}{c^3} - \frac{1.02\,\,10^{-21} \nu^3}{c^4} + \frac{4.08\,\,10^{-27} \nu^4}{c^5} \right],
\end{align}
where Eqs. \ref{Eq-ISRF1} - \ref{Eq-ISRF7} correspond to the cosmic microwave background emission, 
the emissions of the three dust components, the large amorphous carbon and silicates, the small 
amorphous carbon grains, and the PAHs\footnote{for the sake of simplicity, a smooth emission function 
is adopted, neglecting the PAHs band emitters \citep{Draine2007,Compiegne2011}.}, as well as the 
emission of cold, A-type, and OB-type stars. As a reference, the temperatures and opacity laws of 
these functions were chosen to match the interstellar radiation field at high Galactic latitude in 
the ultraviolet \citep{Draine1978,Henry1980}, optical \citep{van-Dishoeck1988,Kopp1996}, near-infrared 
\citep{Moskalenko2006}, mid-infrared \citep{Arendt1998,Compiegne2011}, and far-infrared 
\citep{Lagache1999,Finkbeiner1999} domains. Each component can also be adjusted using the scaling 
factors $\chi_{\rm fir}$, $\chi_{\rm mi_1}$, $\chi_{\rm mi_2}$, $\chi_{\rm nir}$, $\chi_{\rm opt}$, 
and $\chi_{\rm uv}$. The reference parameters are given in Table \ref{Tab-ISRF} and the corresponding 
specific intensity of the local interstellar radiation field is displayed in Fig. \ref{Fig-ISRF}, 
along with the wavelengths of the pure rotational, rovibrational, and rovibronic transitions of \CHp.

Lastly, the absorption of the internal and external radiation fields at near-infrared, optical, and 
ultraviolet wavelengths by dust particles across the cloud is taken into account in MADEX assuming 
the extinction curve $A_\lambda$ derived by \citet{Fitzpatrick1999} and the relation $A_V = 
N_{\rm H}\,\,(\cq) / 1.8 \times 10^{21}$ \citep{Predehl1995} between the visual extinction $A_V$ 
and the total column density $N_{\rm H}$.

\subsection{Analysis of the main excitation pathways}

\begin{figure}[!h]
\begin{center}
\includegraphics[width=9.0cm,angle=0]{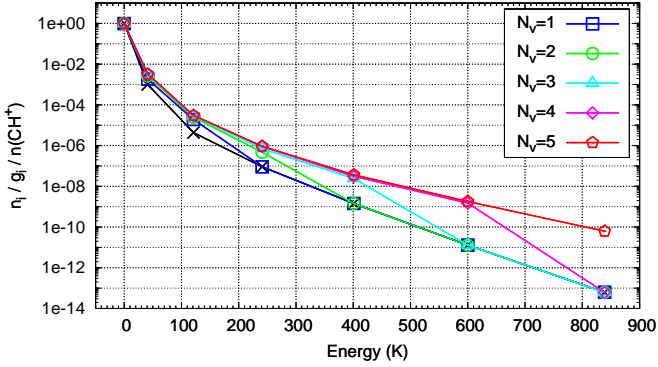}
\caption{\CHp\ steady-state normalized excitation diagrams computed with MADEX assuming 
$\nH = 10^4$ \cc, $\TK = 50$ K, $\chi_{\rm opt} = 10^5$, and $N(\CHp)=10^{13}$ \cq. 
The number of vibrational levels taken into account within the $X^1\Sigma^+$ and $A^1\Pi$ 
electronic states varies between one (blue squares) and five (red pentagons). We also display
for comparison the predictions of MADEX if the radiative pumping processes are switched 
off (black crosses).}
\label{Fig-Cascade}
\end{center}
\end{figure}

Our first goal was to find the optical, near-infrared, and far-infrared radiation fields 
required to perturb the steady-state rotational level populations from what they would be 
in the absence of radiative pumping. Therefore, while more advanced models are discussed 
in the next section, we consider here a spherical cloud of gas only (no extinction from dust) 
without a central star and with constant density, temperature, and chemical composition. 
Moreover, to focus on the radiative pumping of bound rovibronic states, we intentionally 
neglected in a first step the photodissociation process. For $[\HH] = 0.5$, $[\He] = 0.1$, 
$[e^-] = 10^{-4}$, $N(\CHp) = 10^{13}$ \cq, and $\upsilon_{\rm exp} = 1$ \kms, we computed 75600 
models (for a total computing time of $\sim 19$ hours):
\begin{itemize}
\item[$\bullet$] three densities: $\nH =$ $10^3 \times 10^{i-1}$ for $i=1$ to 3;
\item[$\bullet$] 60 kinetic temperatures: $\TK = 30 \times 1.06^{i-1}$ for $i=1$ to 60;
\item[$\bullet$] 140 far-infrared, near-infrared, and optical radiation fields: 
$\chi_{\rm fir}=10^3 \times 1.09^{i-1}$ for $i=1$ to 140, $\chi_{\rm nir}= 1$, and $\chi_{\rm opt}= 1$,
$\chi_{\rm fir}= 1$, $\chi_{\rm nir}=10^6 \times 1.09^{i-1}$ for $i=1$ to 140, and $\chi_{\rm opt}= 1$, and
$\chi_{\rm fir}= 1$, $\chi_{\rm nir}= 1$, and $\chi_{\rm opt}=10^2 \times 1.11^{i-1}$ for $i=1$ to 140, 
\end{itemize}
i.e. exploring the parameter space below the limits set by the critical parameters analysis 
(see Sect. \ref{Section-Exploration} and Appendix \ref{Appen-Critic}). The dominant excitation 
and de-excitation mechanisms are discussed in Appendix \ref{Appen-MainExc} and are shown in Fig. 
\ref{Fig-Main-excitation}. As expected for this parameter range, the de-excitation of the 
rotational levels mainly occurs via the pure rotational radiative transitions for all models. 
Conversely, Fig. \ref{Fig-Main-excitation} reveals five different excitation regimes.
\begin{enumerate}
\item In the limit of low kinetic temperature, the reactive collisions are responsible for the 
excitation of the $J \geqslant 1$ rotational levels, implying $n_{00J} \propto \nH {\rm exp}(-E_{00J}/\TK)$ 
(see Eq. \ref{Eq-sse-che}) at the steady-state equilibrium.
\item In the limit of high kinetic temperature, the $J \geqslant 1$ rotational levels are excited 
through chemical pumping of a high-energy rotational state (e.g. $J=10$) followed by the radiative 
decay of all intermediate $J$ states. Interestingly, the nonreactive collisions contribute 
to less than 20 \% of the rotational excitation even for high values of \TK.
\item In the limit of large far-infrared radiation field, the rotational excitation occurs
through the successive excitations of all intermediate rotational levels by direct
absorption of far-infrared photons. At the steady-state equilibrium, we derive 
$n_{00J} \propto \chi_{\rm fir}^{J}$ (see Eq. \ref{Eq-sse-rot}).
\item In the limit of large near-infrared radiation field, the pure rotational
levels are excited by near-infrared pumping of the first vibrational level of the $X^1\Sigma^+$ 
state. Since the transition dipole moments $M_{0 \upsilon 0 0}$ (for $\upsilon>1$) are at least 
two orders of magnitude smaller than $M_{0 1 0 0}$ (see Table \ref{TabSpec}), the pumping of the 
$\upsilon>1$ levels has a negligible contribution. Moreover, because the transitions are restricted 
to $\Delta J = \pm 1$, the density of \CHp\ in the level $J$ results from the successive excitations 
of all intermediate levels. Consequently, the high-$J$ levels are more difficult to populate 
and are more sensitive to the intensity of the near-infrared radiation field: at steady-state 
$n_{00J} \propto \chi_{\rm nir}^{(J+1)/2}$ and $n_{00J} \propto \chi_{\rm nir}^{J/2}$ (see Eqs. 
\ref{Eq-sse-vib-odd} \& \ref{Eq-sse-vib-even}) for odd and even values of $J$.
\item Finally, in the limit of large optical radiation field, the pure rotational
levels are excited via the optical pumping of the first rotational levels of the $A^1\Pi$ vibronic 
states followed by the radiative decay through the rovibrational levels of the $X^1\Sigma^+$ state.
As shown in Fig. \ref{Fig-Main-excitation}, the selection rules imply that all vibrational 
levels $\upsilon \leqslant J-2$ are required to accurately compute the population of the pure 
rotational level $J$. This result is illustrated in Fig. \ref{Fig-Cascade}, which displays the 
normalized excitation diagrams of \CHp\ (up to $J=6$) computed with MADEX for $\nH=10^4$ \cc, 
$\TK=50$ K, $\chi_{\rm opt}=10^5$ and taking into account an increasing number of vibrational 
levels (from 1 to 5). Lastly, since the excitations of all pure rotational levels ultimately
originate from the same transition $\Lambda,\upsilon,J = 1,\upsilon,1 \leftarrow 0,0,0$, 
their populations are equally sensitive to the intensity of the optical radiation field: $n_{00J} 
\propto \chi_{\rm opt}$ (see Eq. \ref{Eq-sse-ele}) for all $J \geqslant 1$ at the steady-state.
\end{enumerate}

\subsection{Radiation field required to activate the radiative pumping}

\begin{figure}[!h]
\begin{center}
\includegraphics[width=9.0cm,angle=0]{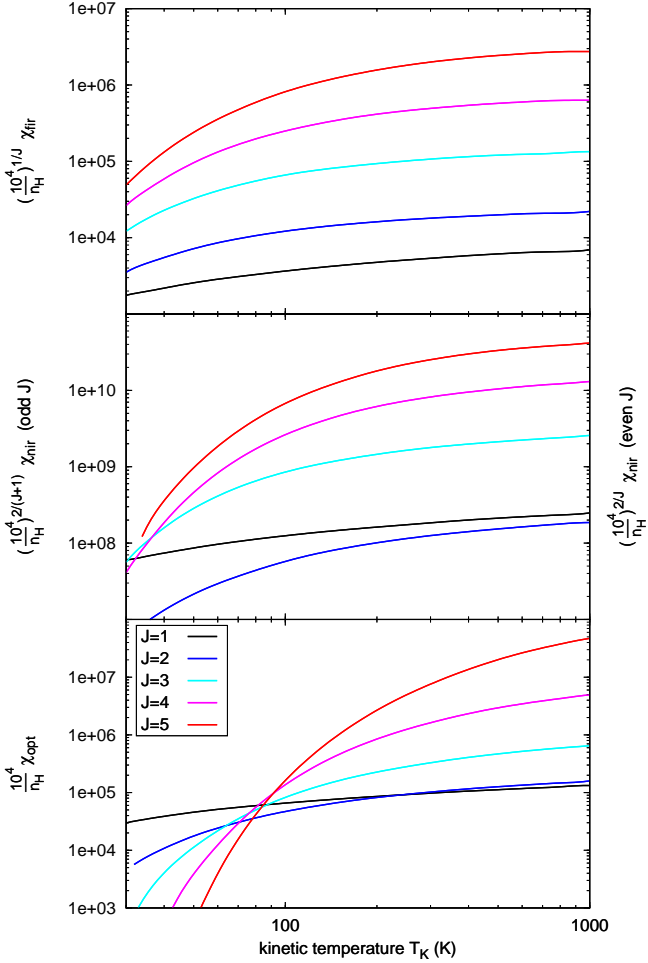}
\caption{Far-infrared (top), near-infrared (middle), and optical (bottom) radiation fields 
(in units of the local ISRF) at which the pure rotational levels of \CHp\ are equally populated 
by collisions and radiative pumping, depending on the kinetic temperature and the density of the 
gas. Below these curves, excitations are primarily driven by reactive collisions and chemical 
pumping. Above these curves, excitations are primarily driven by direct radiative pumping
(top) and by the radiative pumping of the vibrational levels of the $X^1\Sigma^+$ state (middle) 
and of the $A^1\Pi$ electronic state (bottom).}
\label{Fig-Limits}
\end{center}
\end{figure}

With our grid of models we derived the radiation field intensities above which the far-infrared, 
near-infrared, and optical pumpings play a dominant role in populating the excited levels of \CHp. 
These limits are displayed in Fig. \ref{Fig-Limits} for the $J \leqslant 5$ rotational levels as 
functions of the kinetic temperature and the density. The main excitation processes are found to 
exponentially depend on \TK, as was already predicted in the previous section. In addition, the 
superimposition of the curves computed at different densities confirms their dependence on $\nH$ 
and $J$ derived analytically in Appendix \ref{Appen-MainExc}.

Fig. \ref{Fig-Limits} shows that the interaction with the far-infrared radiation competes
with the collisional processes in the rotational excitation of \CHp\ for $\chi_{\rm fir} 
\geqslant 10^4 \times (\nH/10^4)^{1/J}$. Similarly, the pumping of the vibrational states 
could influence the populations of the low-J rotational levels, providing that the gas is 
pervaded by a near-infrared radiation field at least $10^7 \times (\nH/10^4)^{2/J}$ and 
$10^8 \times (\nH/10^4)^{2/J}$ times larger than the local ISRF for $\TK < 100$ K and 
$\TK > 100$ K. Lastly, an optical radiation field of $\sim 10^5 \times (\nH/10^4)$ 
the local ISRF is required to activate the pumping of bound electronic states for $\TK > 
100$ K, while this limit drops at lower kinetic temperatures.

\subsection{Impact of the photodissociation}

As emphasized in Appendix \ref{Appen-Critic}, the impact of the photodissociation on the 
rotational excitation of \CHp\ strongly depends on the spectrum of the UV radiation field. 
For instance, while the photodissociation is negligible compared to bound-bound electronic 
transitions in a radiation field emitted by a black body with an effective temperature 
$< 10^4$ K, the two processes have comparable rates in the standard interstellar radiation 
field. Therefore the photodissociation could play an important role in the rotational 
excitation of \CHp\ in any environment in which the $A^1\Pi$ - $X^1\Sigma^+$ pumping 
competes with the collisional processes (Fig. \ref{Fig-Limits} bottom).

Hereafter, we therefore included the photodissociation process into MADEX and applied a model 
to several astronomical sources with different values of $\TK$, $\chi_{\rm fir}/\nH$, 
$\chi_{\rm nir}/\nH$, and $\chi_{\rm opt}/\nH$. The following questions were addressed: 
how do the predictions of MADEX compare with the available observations? How do all these 
excitation processes modify our understanding of the physical conditions of the interstellar 
and circumstellar matter?

\section{Application to astronomical sources}

\begin{table*}[!ht]
\begin{center}
\caption{Comparison between the observed intensities $\int I d\upsilon$ or fluxes $\iint 
I d\upsilon d\Omega$ of the rotational lines of \CHp\ and those predicted by MADEX for 
the selected astronomical sources. In all cases, MADEX was run in the three configurations 
\ding{182}, \ding{183}, and \ding{184} (see main text), and the far-infrared continuum 
emission was removed in the computation of the line intensities.}
\begin{tabular}{l @{\hspace{1.0cm}} c c c c @{\hspace{1.0cm}} c c c c @{\hspace{1.0cm}} c c c c}
\hline
parameter                   & \multicolumn{4}{c}{cold diffuse medium}                              & \multicolumn{4}{c}{Orion Bar}                              & \multicolumn{4}{c}{NGC 7027} \\
\hline
\nH (\cc)                   & \multicolumn{4}{c}{50}                                               & \multicolumn{4}{c}{$5 \times 10^4$   }                     & \multicolumn{4}{c}{$2 \times 10^5$     } \\
$\chi_{\rm opt}$            & \multicolumn{4}{c}{1 - 50}                                           & \multicolumn{4}{c}{$3 \times 10^4$   }                     & \multicolumn{4}{c}{$4 \times 10^4$     } \\
$\chi_{\rm nir}$            & \multicolumn{4}{c}{1 - 50}                                           & \multicolumn{4}{c}{$3 \times 10^4$   }                     & \multicolumn{4}{c}{$2 \times 10^5$     } \\
$\chi_{\rm fir}$            & \multicolumn{4}{c}{1 - 10}                                           & \multicolumn{4}{c}{$1$               }                     & \multicolumn{4}{c}{$1$                 } \\
$T_{\rm fir}$               & \multicolumn{4}{c}{$18$                }                             & \multicolumn{4}{c}{$55$              }                     & \multicolumn{4}{c}{$150$               } \\
$\tau_{\rm fir}$            & \multicolumn{4}{c}{$1.7 \times 10^{-5}$}                             & \multicolumn{4}{c}{$5 \times 10^{-2}$}                     & \multicolumn{4}{c}{$1 \times 10{-2}$   } \\
$\beta_{\rm fir}$           & \multicolumn{4}{c}{$2.0$               }                             & \multicolumn{4}{c}{$1.75$            }                     & \multicolumn{4}{c}{$0.5$               } \\
$\upsilon_{\rm exp}$ (\kms) & \multicolumn{4}{c}{5}                                                & \multicolumn{4}{c}{10}                                     & \multicolumn{4}{c}{20                  } \\
$\Omega$ (sr)               & \multicolumn{4}{c}{}                                                 & \multicolumn{4}{c}{$2.7 \times 10^{-8}$}                   & \multicolumn{4}{c}{$1.2 \times 10^{-9}$} \\
\hline
transition                  & \multicolumn{4}{c}{intensity}                                        & \multicolumn{4}{c}{flux}                                   & \multicolumn{4}{c}{flux                } \\
                            & \multicolumn{4}{c}{(in $10^{-10}$ erg s$^{-1}$ cm$^{-2}$ sr$^{-1}$)} & \multicolumn{4}{c}{(in $10^{-14}$ erg s$^{-1}$ cm$^{-2}$)} & \multicolumn{4}{c}{(in $10^{-19}$ W cm$^{-2}$)} \\
\hline
                            & \ding{182} & \ding{183} & \ding{184} & obs$^a$ & \ding{182} & \ding{183} & \ding{184} & obs$^b$ & \ding{182} & \ding{183} & \ding{184} & obs$^c$ \\
$J=1-0$ & 2.7 & 4.5 & 5.9 & $< 79$ &  9.3 & 11.8 & 11.8 & $11.1 \pm 1.8$ & 0.4 & 0.5 & 0.5 & $0.47 \pm 0.01$ \\
$J=2-1$ & 3.7 & 7.1 & 9.5 & $<480$ & 13.2 & 18.1 & 18.3 &                & 1.6 & 2.0 & 2.0 & $1.51 \pm 0.05$ \\
$J=3-2$ & 2.4 & 7.0 & 9.7 &        &  6.7 & 13.8 & 13.9 &                & 2.1 & 2.5 & 2.5 & $2.18 \pm 0.17$ \\ 
$J=4-3$ & 2.3 & 7.7 & 1.1 &        &  6.1 & 13.4 & 13.5 &                & 1.6 & 2.1 & 2.1 & $2.00 \pm 0.22$ \\
$J=5-4$ & 1.1 & 6.6 & 9.6 &        &  3.0 &  8.3 &  8.4 &                & 1.1 & 1.7 & 1.7 & $2.50 \pm 0.41$ \\
$J=6-5$ & 1.1 & 6.3 & 9.2 &        &  2.9 &  7.5 &  7.6 &                & 0.7 & 1.3 & 1.3 & $2.41 \pm 0.33$ \\
\hline
\end{tabular}
\begin{list}{}{}
\item[$^a$] detection limits computed with the rms noise level of the \CHp\ $(1-0)$ and $(2-1)$ 
spectra with the Herschel/HIFI instrument (\citealt{Godard2012}; M. Gerin, priv. comm.); $^b$ from 
\citet{Naylor2010} and \citet{Habart2010}; $^c$ from \citet{Cernicharo1997} and \citet{Wesson2010}.
\end{list}
\label{Tab-Resul}
\end{center}
\end{table*}

\subsection{Chemical formation of \CHp}

Because an efficient production pathway is required to balance its rapid destruction by 
hydrogenation and dissociative recombination, \CHp\ is believed to be formed either via 
\begin{equation}
\Cpp + \HH \rightarrow \CHp + \Hp
\end{equation}
in environments submitted to a strong X-ray radiation field (e.g. the central molecular zone, 
\citealt{Langer1978}), or by the highly endothermic reaction
\begin{equation} \label{Eq-CHp-form}
\Cp + \HH \rightarrow \CHp + {\rm H} \qquad (\Delta E/k = 4640 K)
\end{equation}
in other Galactic environments. So far, two main scenarios have been invoked to overcome the 
endothermicity of reaction \ref{Eq-CHp-form}: (1) by the release of supra-thermal energy, 
induced for example by magnetohydrodynamic shocks, in regions with large reservoirs of kinetic
and magnetic energies (e.g. \citealt{Draine1986,Xie1995,Falgarone1995}); and (2) by the internal 
energy of \HH\ \citep{Hierl1997,Agundez2010} in regions where the vibrational levels of \HH\ are 
excited by FUV fluorescence.

For each astronomical source considered below, we therefore applied an appropriate chemical 
model to describe the abundance distributions of \CHp\ and of its main collisional partners.
To investigate the origine of the rotational excitation of \CHp\, we then systematically
ran MADEX in three different configurations:
\ding{182} considering only the excitation by nonreactive collisions, 
\ding{183} considering all collisional excitation processes (i.e. including the chemical pumping), and
\ding{184} considering all excitation processes (i.e. including the radiative pumping and the photodissociation).
The parameters and results of all models are summarized in Table \ref{Tab-Resul},
where we compare the continuum-substracted line fluxes (or continuum-substracted line 
intensities) of \CHp\ computed with MADEX with the available observations.

\subsection{Diffuse interstellar medium}

We first focus on the cold diffuse interstellar matter characterized by a low density and by low 
$\chi_{\rm nir}/\nH$ and $\chi_{\rm opt}/\nH$ ratios. The observed mean column density of \CHp, 
$\log_{10}(N(\CHp)/N_{\rm H}) \sim -8.11 \pm 0.33$ (from a compilation of data: \citealt{Crane1995,
Gredel1997,Weselak2008, Falgarone2010,Godard2012}) has been proposed to trace low-velocity 
magnetohydrodynamic shocks \citep{Pineau-des-Forets1986}, Alfvén waves \citep{Federman1996}, 
turbulent mixing \citep{Lesaffre2007}, or turbulent dissipation \citep{Joulain1998}. According
to the TDR (turbulent dissipation region) model of \citet{Godard2009}, \CHp\ is formed in regions
where turbulent dissipation is triggered by the decoupling of the ionized and neutral flows
at small spatial scales ($\sim 100$ AU). In these regions, the kinetic temperature increases to 
$\sim 1000$ K, while the \CHp\ formation temperature is boosted by the ion-neutral velocity drift 
up to $\sim 2000$ K. Assuming a density $\nH = 50$ \cc\ and a moderate shielding $A_V > 0.2$ of the 
interstellar radiation field ($\chi_{\rm opt} = \chi_{\rm nir} = \chi_{\rm fir} = 1$), most 
of the hydrogen is in molecular form and the ionization degree is $\sim 10^{-4}$.

These properties were injected into MADEX to compute the beam-averaged 
continuum-substracted intensities of the \CHp\ rotational lines (see Table \ref{Tab-Resul}) 
assuming that the turbulent dissipation regions have a surface filling factor of 1.
We find that the pumping by near- and far-infrared radiation has no influence on the 
excitation of the rotational levels of \CHp\ even if $\chi_{\rm nir} = 50$ and $\chi_{\rm fir} 
= 10$, i.e. adopting the highest energy density of the interstellar radiation field derived by 
\citet{Moskalenko2006} across the Galactic disk. In contrast, the optical pumping and the
photodissociation are found to be non-negligible compared to the collisional processes 
for $\chi_{\rm uv/opt} \geqslant 10$, i.e. in the most central regions ($R_g < 4$ Kpc, 
\citealt{Moskalenko2006}) of the Milky Way. For instance, for $\chi_{\rm uv/opt} = 50$, the 
optical pumping and the photodissociation contribute up to 30 \% of the rotational excitation
of \CHp (Col. 4 of Table \ref{Tab-Resul}). In all cases, we finally find that the intensities 
of the two first rotational emission lines predicted by MADEX are at least 14 times smaller 
than the limit of detection of the Herschel/HIFI data \citep{Godard2012}, in agreement with 
the fact that \CHp\ has only been detected in the diffuse interstellar medium in absorption 
towards strong submillimeter continuum sources.

\subsection{Hot and dense PDRs: e.g. the Orion Bar}

\begin{figure}[!h]
\begin{center}
\includegraphics[width=9.0cm,angle=0]{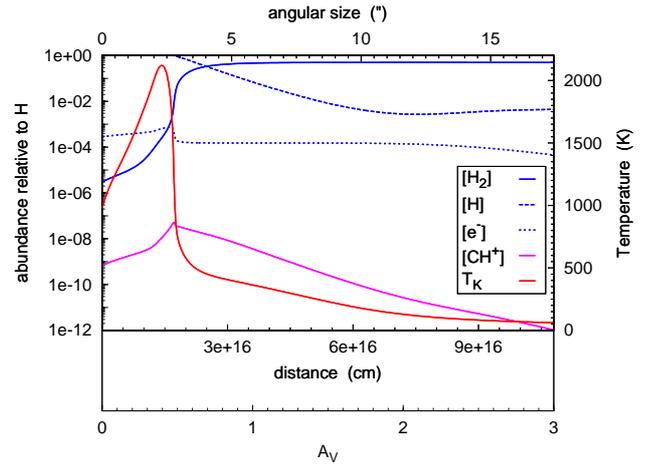}
\caption{Kinetic temperature and abundances relative to H of \HH, H, $e^-$, and 
\CHp\ computed with the Meudon PDR code as functions of the extinction (bottom axis), 
the distance (middle axis), and the angular size (top axis) across the Orion Bar 
starting from the ionization front.}
\label{Fig-Orion}
\end{center}
\end{figure}

We selected the Orion Bar as a prototypical hot and dense PDR with an high $\chi_{\rm opt}/\nH$ ratio. 
Lying at a distance of $414 \pm 7$ pc \citep{Menten2007}, the Bar consists of a homogeneous interclump 
medium with a density $\nH \sim 5 \times 10^4$ \cc\ \citep{Young-Owl2000}, located approximately $111
''$ ($\sim 0.22$ pc) south-east of the Trapezium star cluster \citep{Pellegrini2009}. The resulting 
incident UV and optical radiation fields at the ionization front are $\sim 3 \times 10^4$ that of the 
local ISRF \citep{Marconi1998}. The near-infrared continuum emission of the HII region corresponds 
to $10^4$ that of the local ISRF \citep{Walmsley2000}. Lastly, the recent observations at 70, 160, 
250, 350, and 500 $\mu$m performed by \citet{Arab2012} using the PACS and SPIRE instruments of the 
Herschel space telescope have revealed dust temperature and spectral index gradients from 80 K to 
40 K and from 1.2 to 2.2 from the ionization front to the Orion molecular cloud, respectively.

With all these parameters, we computed the chemical profile of the Orion Bar with the Meudon PDR 
code, a one-dimensional chemical model in which a static slab of gas of given thickness is illuminated 
on one side or on both sides by a given FUV radiation field \citep{Le-Petit2006}. Fig. \ref{Fig-Orion} 
displays the physical structure of the Orion Bar predicted by the most recent online version of the 
Meudon PDR code\footnote{version 1.4.4 available at http://pdr.obspm.fr/PDRcode.html}, which takes 
into account the influence of vibrationally excited \HH\ in the formation of \CHp\ \citep{Agundez2010}. 
Inserting the density profiles of H, \HH, $e^-$, and \CHp\ into MADEX and taking into account the 
geometry of the bar described by \citet{Hogerheijde1995}, we finally derived the six first rotational 
line fluxes of \CHp\ (integrated over a $36''$ wide solid angle, after having removed the 
far-infrared continuum emission) given in columns 6, 7, and 8 of Table \ref{Tab-Resul}.

The comparison of models \ding{182}, \ding{183}, and \ding{184} shows that the radiative pumping
and the photodissociation have only a marginal effect ($\leqslant 2$ \%) on the excitation of the 
rotational levels of \CHp.
In contrast, even if the high-$J$ transitions appear to be mostly driven by chemical pumping, both
reactive and nonreactive collisions are required to accurately model all the far-infrared emission 
lines of \CHp. At last, the flux of the \CHp\ $J=1-0$ transition derived with MADEX is found to 
agree well with the value observed by \citet{Naylor2010} and \citet{Habart2010} using the Herschel/SPIRE 
instrument. This result indicates that the emission lines of \CHp\ detected in the Orion Bar probably 
arise from the interclump medium rather than from dense and hot regions ($\TK \sim 200$ K and 
$\nH \sim 10^7$ \cc, \citealt{Goicoechea2011}). Since the $J=3-2$ transition of \CHp\ detected with
Herschel/PACS was found to spatially correlate with the 60-100 $\mu$m rotational lines of OH
\citep{Goicoechea2011} and the high-$J$ rotational lines of CO detected with Herschel/SPIRE 
\citep{Habart2010}, this result also suggests that chemical formation and/or radiative pumping 
could have a significant impact on the rotational excitation of OH and CO in the Orion Bar PDR. 
If this is the case, clumps would no longer be required to account for their observed emission 
line intensities \citep{Goicoechea2011}.

\subsection{Planetary nebulae: e.g. NGC 7027}

\begin{figure}[!h]
\begin{center}
\includegraphics[width=9.0cm,angle=0]{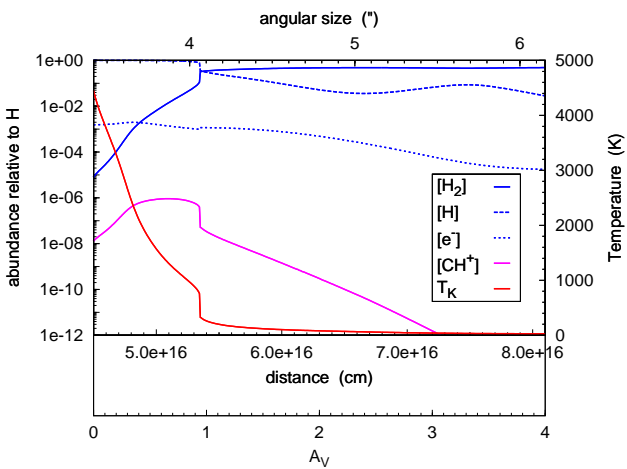}
\caption{Kinetic temperature and abundances relative to H of \HH, H, $e^-$, and \CHp\ computed with 
the Meudon PDR code as functions of the extinction (bottom axis) across the PDR region of NGC 7027
starting from the ionization front. The distance and the angle from the central star are indicated
along the middle and top axes.}
\label{Fig-NGC7027}
\end{center}
\end{figure}

Lastly, we chose NGC 7027 as an archetypal planetary nebula with an high $\chi_{\rm nir}/\nH$ ratio.
Located at a distance of $880 \pm 150$ pc \citep{Masson1989}, NGC 7027 is composed of ionized, 
atomic, and molecular C-rich shells surrounding a single central star with an effective temperature 
$T_{\star} \sim 198,000$ K and a radius $R_{\star} \sim 5.21 \times 10^9$ cm \citep{Latter2000}. 
The ionized shell revealed by a strong near-infrared continuum emission has an ellipsoidal 
morphology with dimensions $6'' \times 12''$, while the borders of the PDR traced by the emission 
lines of \HH\ have a biconal structure with dimensions $10'' \times 13''$ \citep{Latter2000,Cox2002}. 
The molecular envelope detected in CO and other millimetric molecular transitions extends up to
$\sim 100''$ in radius \citep{Navarro2003}.

For the sake of simplicity, we modelled the PDR region as a spherical shell with dimensions set 
to match the molecular emission lines observed along the minor axis of NGC 7027 \citep{Cox2002}. 
The ionization front is located at $3.4''$ ($\sim 0.0145$ pc) from the central star, implying 
an incident UV radiation field of $\sim 3 \times 10^4$ that of the local ISRF \citep{Hasegawa2000}.
Using the optical and near-infrared spectra modelled by \citet{Volk1997} and the extinction curve
found by \citet{Zhang2003}, we estimate that the incident optical and near-infrared radiation 
fields are $\sim 4 \times 10^4$ and $\sim 2 \times 10^5$ times stronger than the local ISRF at 
0.4 $\mu$m and 4 $\mu$m. Finally, the far- and mid-infrared dust emission are modelled as a 
single grey body with a temperature $T_{\rm fir} = 150$ K, an opacity $\tau_{\rm fir} 
= 10^{-5}$, and a spectral index $\beta_{\rm fir} = 0.5$ to reproduce the continuum
emission observed by \citet{Dyck1976} and \citet{Terzian1989}. In contrast to the work of 
previous authors \citep{Cox2002,Hasegawa2000,Yan1999}, who proposed PDR models including a 
high-density shell, we considered only the low-density PDR component defined by $\nH = 2 
\times 10^5$ \cc.

In the same way as for the Orion Bar PDR, the chemical composition of NGC 7027 was computed with
the Meudon PDR code, i.e. assuming that the radial structure of the source is well described 
with a plan-parallel model. The carbon-rich circumstellar envelope was modelled with C, O, N, 
and S elemental abundances of $1.3 \times 10^{-3}$, $5.5 \times 10^{-4}$, $1.9 \times 10^{-4}$, 
and $7.9 \times 10^{-6}$ \citep{Middlemass1990}. The resulting chemical and temperature profiles 
are shown in Fig. \ref{Fig-NGC7027} as functions of the distance from the central star. The fluxes 
of the rotational emission lines of \CHp\ (integrated over a $9''$ wide solid angle, after removing 
the far-infrared continuum emission, see Table \ref{Tab-Resul}) were finally calculated, assuming 
that the chemical composition has a spherical symmetry and taking into account the influence of the 
limb brightening.

Because of the high kinetic temperature and electronic fraction, the nonreactive collisions of 
\CHp\ with $e^-$ pilot the distribution of \CHp\ among the low-$J$ rotational levels, and even 
compete with the chemical pumping in the excitation of the high-$J$ transitions. Our main result 
is that the predictions of MADEX agree excellently with the observations performed by 
\citet{Cernicharo1997} using the ISO/SWS instrument and with those of \citet{Wesson2010}
using the Herschel/SPIRE instrument. Furthermore, because all collisional excitation 
processes are included in MADEX, the intensities of all observed rotational lines are 
explained with a circumstellar medium of density $\nH = 2 \times 10^5$, a value 10 to 100 
times lower than those inferred from traditional excitation models \citep{Cernicharo1997,
Hasegawa2000,Yan1999}.

\subsection{Summary} \label{Summary}

We conclude from these applications that the far-infrared pumping, the near-infrared 
and optical fluorescences, and the photodissociation have a marginal impact on the 
far-infrared emission lines of \CHp. For all the sources considered,
the excitations of the high-$J$ and low-$J$ lines mainly arise from chemical pumping and from 
a combination of chemical pumping and nonreactive collisional processes. For 
low-$J$ lines, the balance of these two mechanisms then principally depends on the chemical 
composition of the gas and the kinetic temperature. For instance, in environments with a large 
electronic fraction, such as the C-rich PDRs (e.g. NGC 7027), the nonreactive collisions with 
electrons are found to dominate the rotational excitation of \CHp\ up to the $J=4-3$ transition.

Our main result is that the comparison between the observed line fluxes and the predictions
of MADEX points towards interstellar and circumstellar media with densities far lower than 
those deduced with traditional models. This study thus proves the importance of exploring all 
possible excitation pathways of a molecule before drawing any conclusion on the physical 
structure and geometry of the source. In particular, we question the degree of clumpiness of 
the Orion Bar molecular cloud. 

We therefore recommend to treat any other astronomical environments detected in \CHp in a 
similar way. For instance, the rotational lines of \CHp\ have recently been observed by 
\citet{Thi2011} in the protoplanetary disc that surrounds the Herbig Be star HD 100546 
using the Herschel/PACS instrument. Considering only the excitation by nonreactive collisions, 
\citet{Thi2011} found that the strong 
emission lines of \CHp\ (up to $J=6-5$) arise from the rim of the outer disc and the disc 
surface, i.e. from media predominantly atomic with densities $\sim 10^8$ \cc\ and temperatures
$\sim 10^{2} - 10^{3}$ K. With an effective temperature of the central star $\sim 10 500$ K
\citep{van-den-Ancker1997}, the photodissociation is unlikely to have any effect on the 
distribution of \CHp\ among its rotational levels (see Appendix \ref{Appen-Critic}).
Conversely, the strong flux of optical photons could enhance the high-$J$ level populations
at the disc surface where the \CHp\ abundance is the highest. Moreover, with the chemical 
composition described by \citet{Thi2011}, we expect the chemical pumping to dominate
the nonreactive collisional processes. It would thus be interesting to estimate how
the chemical and radiative pumping affect the understanding of the density structure of 
the disc and of the spatial emission of rotationally excited \CHp.

In preparation of future investigations, we present below an extension of our model to other 
diatomic molecules with spectroscopic characteristics similar to those of \CHp.

\section{Discussion: extension to other diatomic molecules} \label{Discussion}

\begin{table*}[!ht]
\begin{center}
\caption{Predictions of MADEX applied to \CHp\ and SiO.}
\begin{tabular}{c @{\hspace{0.2cm}} c @{\hspace{-0.0cm}} c | c @{\hspace{0.2cm}} c @{\hspace{0.2cm}} c @{\hspace{0.2cm}} c @{\hspace{0.2cm}} c @{\hspace{0.2cm}} c | c @{\hspace{0.2cm}} c @{\hspace{0.2cm}} c @{\hspace{0.2cm}} c @{\hspace{0.2cm}} c @{\hspace{0.2cm}} c @{\hspace{0.2cm}}}
\hline
& &                   & \multicolumn{6}{c|}{$^a$\CHp}                                                                                               & \multicolumn{6}{c}{$^b$SiO}                                                               \\
\hline
& & & & & & & & & & & & \\
& & $\Lambda$         & \multicolumn{2}{c}{$T_e$ (cm$^{-1}$)} & \multicolumn{2}{c}{$\omega_e$ (cm$^{-1}$)} & \multicolumn{2}{c|}{$B_e$ (cm$^{-1}$)} & \multicolumn{2}{c}{$T_e$ (cm$^{-1}$)} & \multicolumn{2}{c}{$\omega_e$ (cm$^{-1}$)} & \multicolumn{2}{c}{$B_e$ (cm$^{-1}$)} \\[4pt]
& & $0$               & \multicolumn{2}{c}{}                  & \multicolumn{2}{c}{2.858 ($+03$)}          & \multicolumn{2}{c|}{1.418 ($+01$)}     & \multicolumn{2}{c}{}                  & \multicolumn{2}{c}{1.242 ($+03$)}          & \multicolumn{2}{c}{7.242 ($-01$)}     \\
& & $1$               & \multicolumn{2}{c}{2.412 ($+04$)}     & \multicolumn{2}{c}{1.864 ($+03$)}          & \multicolumn{2}{c|}{1.189 ($+01$)}     & \multicolumn{2}{c}{4.284 ($+04$)}     & \multicolumn{2}{c}{8.528 ($+02$)}          & \multicolumn{2}{c}{6.307 ($-01$)}     \\
& & & & & & & & & & & & \\
\hline
& & & & & & & & & & & & \\
& &                   & \multicolumn{6}{c|}{$M_{\Lambda'\upsilon'\Lambda''\upsilon''}$ ($ea_0$)} & \multicolumn{6}{c}{$M_{\Lambda'\upsilon'\Lambda''\upsilon''}$ ($ea_0$)}  \\[3pt]
& & $\upsilon''$\textbackslash$\upsilon'$
                      & $0$        & $1$        & $2$        & $3$        & $4$        &                & $0$        & $1$        & $2$        & $3$        & $4$        &                \\[3pt]
\multirow{5}{*}{\begin{sideways}$\Lambda' \rightarrow \Lambda''$ \end{sideways}} & \multirow{5}{*}{\begin{sideways}$1 \rightarrow 0$ \end{sideways}}
  & $0$               & 2.2 ($-1$) & 1.6 ($-1$) & 1.0 ($-1$) & 6.5 ($-2$) & 4.1 ($-2$) &                & 2.0 ($+0$) & 2.7 ($+0$) & 2.6 ($+0$) & 2.0 ($+0$) & 1.3 ($+0$) &                \\
& & $1$               & 1.1 ($-1$) & 9.1 ($-2$) & 1.5 ($-1$) & 1.3 ($-1$) & 1.1 ($-1$) &                & 2.7 ($+0$) & 1.9 ($+0$) & 1.6 ($+0$) & 1.5 ($+0$) & 2.1 ($+0$) &                \\
& & $2$               & 4.6 ($-2$) & 1.1 ($-1$) & 1.6 ($-2$) & 7.0 ($-2$) & 1.0 ($-1$) &                & 2.8 ($+0$) & 1.4 ($-1$) & 1.8 ($+0$) & 1.5 ($+0$) & 6.6 ($-2$) &                \\
& & $3$               & 1.6 ($-2$) & 7.2 ($-2$) & 6.5 ($-2$) & 6.7 ($-2$) & 1.5 ($-2$) &                & 2.4 ($+0$) & 1.3 ($+0$) & 1.7 ($+0$) & 3.4 ($-1$) & 1.6 ($+0$) &                \\
& & $4$               & 5.1 ($-3$) & 3.5 ($-2$) & 7.6 ($-2$) & 2.1 ($-3$) & 5.2 ($-2$) &                & 1.8 ($+0$) & 2.1 ($+0$) & 4.8 ($-1$) & 1.6 ($+0$) & 9.8 ($-1$) &                \\
& & & & & & & & & & & & \\
& &                   & \multicolumn{6}{c|}{$M_{\Lambda'\upsilon'\Lambda''\upsilon''}$ ($ea_0$)} & \multicolumn{6}{c}{$M_{\Lambda'\upsilon'\Lambda''\upsilon''}$ ($ea_0$)}  \\[3pt]
& & $\upsilon''$\textbackslash$\upsilon'$
                      & $0$        & $1$        & $2$        & $3$        & $4$        &                & $0$        & $1$        & $2$        & $3$        & $4$        &                \\[3pt]
\multirow{5}{*}{\begin{sideways}$\Lambda' \rightarrow \Lambda''$ \end{sideways}} & \multirow{5}{*}{\begin{sideways}$0 \rightarrow 0$ \end{sideways}}
  & $0$               & 6.6 ($-1$) & 6.3 ($-3$) & 6.6 ($-5$) & 6.5 ($-7$) & 6.5 ($-9$) &                & 2.0 ($+0$) & 4.1 ($-2$) & 2.3 ($-3$) & 1.6 ($-4$) & 1.9 ($-5$) &                \\
& & $1$               &            & 6.6 ($-1$) & 9.6 ($-3$) & 7.7 ($-4$) & 2.0 ($-5$) &                &            & 6.1 ($+0$) & 5.7 ($-2$) & 4.4 ($-3$) & 3.2 ($-4$) &                \\
& & $2$               &            &            & 6.7 ($-1$) & 1.3 ($-2$) & 2.9 ($-3$) &                &            &            & 1.2 ($+1$) & 6.9 ($-2$) & 6.2 ($-3$) &                \\
& & $3$               &            &            &            & 6.7 ($-1$) & 1.8 ($-2$) &                &            &            &            & 1.8 ($+1$) & 7.8 ($-2$) &                \\
& & $4$               &            &            &            &            & 6.8 ($-1$) &                &            &            &            &            & 2.0 ($+1$) &                \\
& & & & & & & & & & & & \\
\hline
& & & & & & & & & & & & \\
& &                   & \multicolumn{6}{c|}{$\sum\,[C]\,k^C_{J'J''}$ (cm$^3$ s$^{-1}$)}            & \multicolumn{6}{c}{$\sum\,[C]\,k^C_{J'J''}$ (cm$^3$ s$^{-1}$)}             \\[3pt]
& & $J''$\textbackslash$J'$
                      & $0$       & $1$       & $2$       & $3$       & $4$       & $5$            & $0$       & $1$       & $2$       & $3$       & $4$       & $5$            \\[3pt]
& & $0$               &           & 8.1 (-11) & 5.3 (-11) & 1.1 (-11) & 8.4 (-12) & 9.1 (-13)      &           & 2.8 (-11) & 1.1 (-11) & 7.3 (-12) & 1.2 (-12) & 2.1 (-12)      \\
& & $1$               & 1.6 (-10) &           & 8.8 (-11) & 7.5 (-11) & 1.2 (-11) & 1.9 (-11)      & 8.4 (-11) &           & 4.9 (-11) & 1.6 (-11) & 1.4 (-11) & 2.0 (-12)      \\
& & $2$               & 8.0 (-11) & 6.6 (-11) &           & 5.9 (-11) & 8.4 (-11) & 5.4 (-12)      & 5.0 (-11) & 7.8 (-11) &           & 5.4 (-11) & 1.9 (-11) & 1.7 (-11)      \\
& & $3$               & 7.0 (-12) & 2.4 (-11) & 2.5 (-11) &           & 3.7 (-11) & 7.1 (-11)      & 4.5 (-11) & 3.4 (-11) & 7.1 (-11) &           & 5.5 (-11) & 2.0 (-11)      \\
& & $4$               & 1.4 (-12) & 9.5 (-13) & 9.2 (-12) & 9.6 (-12) &           & 2.4 (-11)      & 8.5 (-12) & 3.4 (-11) & 2.9 (-11) & 6.6 (-11) &           & 5.6 (-11)      \\
& & $5$               & 2.5 (-14) & 2.5 (-13) & 9.8 (-14) & 3.1 (-12) & 4.0 (-12) &                & 1.7 (-11) & 5.5 (-12) & 2.8 (-11) & 2.6 (-11) & 6.1 (-11) &                \\
& & & & & & & & & & & & \\
\hline
& & & & & & & & & & & & \\
& &                   & \multicolumn{3}{c|}{$\sum\,[D]\,k^D$ (cm$^3$ s$^{-1}$)} & \multicolumn{3}{c|}{$\int \frac{4\pi}{h\nu} \sigma^\gamma(\nu) I_\nu d\nu$ (s$^{-1}$)} & \multicolumn{3}{c|}{$\sum\,[D]\,k^D$ (cm$^3$ s$^{-1}$)} & \multicolumn{3}{c}{$\int \frac{4\pi}{h\nu} \sigma^\gamma(\nu) I_\nu d\nu$ (s$^{-1}$)} \\[3pt]
& &                   & \multicolumn{3}{c|}{6.24 ($-10$)}                       & \multicolumn{3}{c|}{3.30 ($-10$)}                                                      & \multicolumn{3}{c|}{9.35 ($-14$)}                       & \multicolumn{3}{c}{1.60 ($-09$)}                                                       \\
& & & & & & & & & & & & \\
\hline
& & & & & & & & & & & & \\
& &                   & \multicolumn{6}{c|}{$n_J/n(\CHp)$}                                         & \multicolumn{6}{c}{$n_J/n({\rm SiO})$}                                     \\[3pt]
& & $J$               & $0$       & $1$       & $2$       & $3$       & $4$       & $5$            & $0$       & $1$       & $2$       & $3$       & $4$       & $5$            \\[3pt]
& & \ding{182}        & 1.0 (+00) & 1.4 (-03) & 1.5 (-05) & 3.8 (-07) & 2.6 (-08) & 3.2 (-10)      & 2.6 (-01) & 4.5 (-01) & 2.3 (-01) & 4.2 (-02) & 3.5 (-03) & 3.1 (-04)      \\[3pt]
& & \ding{183}        & 1.0 (+00) & 4.3 (-03) & 5.9 (-05) & 4.8 (-06) & 4.5 (-07) & 3.4 (-08)      & 2.6 (-01) & 4.5 (-01) & 2.3 (-01) & 4.2 (-02) & 3.5 (-03) & 3.1 (-04)      \\[3pt]
\multirow{3}{*}{\begin{sideways}$\chi_{\rm nir}$   \end{sideways}} 
& $10^3$ & \ding{184} & 1.0 (+00) & 4.3 (-03) & 5.9 (-05) & 4.8 (-06) & 4.5 (-07) & 3.4 (-08)      & 2.6 (-01) & 4.5 (-01) & 2.3 (-01) & 4.2 (-02) & 3.5 (-03) & 3.1 (-04)      \\
& $10^5$ & \ding{184} & 1.0 (+00) & 4.3 (-03) & 5.9 (-05) & 4.8 (-06) & 4.5 (-07) & 3.4 (-08)      & 2.6 (-01) & 4.5 (-01) & 2.3 (-01) & 4.2 (-02) & 3.6 (-03) & 3.1 (-04)      \\
& $10^7$ & \ding{184} & 1.0 (+00) & 4.6 (-03) & 7.0 (-05) & 4.8 (-06) & 4.5 (-07) & 3.4 (-08)      & 2.1 (-01) & 4.2 (-01) & 2.8 (-01) & 7.1 (-02) & 7.2 (-03) & 5.4 (-04)      \\[3pt]
\multirow{3}{*}{\begin{sideways}$\chi_{\rm opt/UV}$\end{sideways}} 
& $10^2$ & \ding{184} & 1.0 (+00) & 4.3 (-03) & 5.9 (-05) & 4.8 (-06) & 4.5 (-07) & 3.4 (-08)      & 2.5 (-01) & 4.5 (-01) & 2.3 (-01) & 4.5 (-02) & 4.0 (-03) & 3.7 (-04)      \\
& $10^3$ & \ding{184} & 1.0 (+00) & 4.3 (-03) & 6.1 (-05) & 4.9 (-06) & 4.5 (-07) & 3.4 (-08)      & 2.1 (-01) & 4.3 (-01) & 2.6 (-01) & 7.1 (-02) & 9.3 (-03) & 1.1 (-03)      \\
& $10^4$ & \ding{184} & 9.9 (-01) & 5.0 (-03) & 7.2 (-05) & 5.4 (-06) & 4.8 (-07) & 3.6 (-08)      & 9.3 (-02) & 3.0 (-01) & 3.1 (-01) & 1.9 (-01) & 7.1 (-02) & 1.8 (-02)      \\[3pt]
\multirow{3}{*}{\begin{sideways}$\chi_{\rm opt/UV}$\end{sideways}} 
& $10^2$ & \ding{185} & 1.0 (+00) & 4.3 (-03) & 6.0 (-05) & 4.8 (-06) & 4.5 (-07) & 3.4 (-08)      & 2.5 (-01) & 4.5 (-01) & 2.3 (-01) & 4.8 (-02) & 4.5 (-03) & 4.8 (-04)      \\
& $10^3$ & \ding{185} & 1.0 (+00) & 4.5 (-03) & 6.3 (-05) & 5.1 (-06) & 4.7 (-07) & 3.6 (-08)      & 1.9 (-01) & 4.1 (-01) & 2.8 (-01) & 9.0 (-02) & 1.5 (-02) & 2.4 (-03)      \\
& $10^4$ & \ding{185} & 9.9 (-01) & 6.4 (-03) & 9.9 (-05) & 8.0 (-06) & 7.2 (-07) & 5.5 (-08)      & 6.6 (-02) & 2.5 (-01) & 2.9 (-01) & 2.2 (-01) & 1.1 (-01) & 3.8 (-02)      \\
& & & & & & & & & & & & \\
\hline
\end{tabular}
\begin{list}{}{}
\item[$$] From top to bottom: molecular constants of the $X^1\Sigma^+$ and $A^1\Pi$ electronic 
states, transition dipole moments of the  $A^1\Pi - X^1\Sigma^+$ and $X^1\Sigma^+ - X^1\Sigma^+$ 
band systems, nonreactive collisional rates, chemical destruction rates and photodissociation rates, 
and populations of the rotational levels predicted with MADEX for $J \leqslant 5$. The collisional 
rates and the predictions of the level populations are computed for $\TD=\TF=\TK=100$ K, $\nH=10^4$ 
\cc, $[\HH]=0.5$, and $[e^-]=10^{-4}$. For both molecules, MADEX was run in the four configurations 
\ding{182}, \ding{183}, \ding{184}, and \ding{185} (see main text). In cases \ding{184} and \ding{185}, 
the values of the radiation field scaling factors $\chi_{\rm nir}$ and $\chi_{\rm opt/UV}$ are 1 unless 
specified otherwise. Numbers in parenthesis are power of ten.
\item[$^a$]data from \citet{Hakalla2006,Cheng2007,Lim1999,Hammami2008,Turpin2010}; $^b$data from 
\citet{Hedelund1972,Beer1974,Barrow1975,Elander1973,Naidu1981,Drira1997,Tipping1981,Liszt1972,Drira1998,Park1993,Dayou2006}
\end{list}
\label{Tab-Extension1}
\end{center}
\end{table*}

\begin{table*}[!ht]
\begin{center}
\caption{Same as Table \ref{Tab-Extension1} for HF and CS.}
\begin{tabular}{c @{\hspace{0.2cm}} c @{\hspace{-0.0cm}} c | c @{\hspace{0.2cm}} c @{\hspace{0.2cm}} c @{\hspace{0.2cm}} c @{\hspace{0.2cm}} c @{\hspace{0.2cm}} c | c @{\hspace{0.2cm}} c @{\hspace{0.2cm}} c @{\hspace{0.2cm}} c @{\hspace{0.2cm}} c @{\hspace{0.2cm}} c @{\hspace{0.2cm}}}
\hline
& &                   & \multicolumn{6}{c|}{$^a$HF}                                                                                                 & \multicolumn{6}{c}{$^b$CS}                                                                                                 \\
\hline
& & & & & & & & & & & & \\
& & $\Lambda$         & \multicolumn{2}{c}{$T_e$ (cm$^{-1}$)} & \multicolumn{2}{c}{$\omega_e$ (cm$^{-1}$)} & \multicolumn{2}{c|}{$B_e$ (cm$^{-1}$)} & \multicolumn{2}{c}{$T_e$ (cm$^{-1}$)} & \multicolumn{2}{c}{$\omega_e$ (cm$^{-1}$)} & \multicolumn{2}{c}{$B_e$ (cm$^{-1}$)} \\[3pt]
& & $0$               & \multicolumn{2}{c}{}                  & \multicolumn{2}{c}{4.138 ($+03$)}          & \multicolumn{2}{c|}{2.056 ($+01$)}     & \multicolumn{2}{c}{}                  & \multicolumn{2}{c}{1.285 ($+03$)}          & \multicolumn{2}{c}{8.200 ($-01$)}     \\
& & $1$               & \multicolumn{2}{c}{}                  & \multicolumn{2}{c}{}                       & \multicolumn{2}{c|}{}                  & \multicolumn{2}{c}{3.890 ($+04$)}     & \multicolumn{2}{c}{1.073 ($+03$)}          & \multicolumn{2}{c}{7.800 ($-01$)}     \\
& & & & & & & & & & & & \\
\hline
& & & & & & & & & & & & \\
& &                   & \multicolumn{6}{c|}{$M_{\Lambda'\upsilon'\Lambda''\upsilon''}$ ($ea_0$)} & \multicolumn{6}{c}{$M_{\Lambda'\upsilon'\Lambda''\upsilon''}$ ($ea_0$)}  \\[3pt]
& & $\upsilon''$\textbackslash$\upsilon'$
                      & $0$       & $1$       & $2$       & $3$       & $4$       &                & $0$       & $1$       & $2$       & $3$       & $4$       &                \\[3pt]
\multirow{5}{*}{\begin{sideways}$\Lambda' \rightarrow \Lambda''$ \end{sideways}} & \multirow{5}{*}{\begin{sideways}$1 \rightarrow 0$ \end{sideways}}
  & $0$               &           &           &           &           &           &                & 2.0 ($-1$) & 9.2 ($-2$) & 2.9 ($-2$) & 6.6 ($-3$) &            &                \\
& & $1$               &           &           &           &           &           &                & 6.4 ($-2$) & 1.6 ($-1$) & 1.2 ($-1$) & 5.2 ($-2$) & 3.6 ($-2$) &                \\
& & $2$               &           &           &           &           &           &                & 1.9 ($-2$) & 8.3 ($-2$) & 1.2 ($-1$) & 1.3 ($-1$) & 2.2 ($-1$) &                \\
& & $3$               &           &           &           &           &           &                & 4.8 ($-3$) & 3.3 ($-2$) & 9.1 ($-2$) & 8.3 ($-2$) & 4.7 ($-1$) &                \\
& & $4$               &           &           &           &           &           &                & 7.8 ($-3$) & 6.2 ($-2$) & 2.4 ($-1$) & 4.2 ($-1$) & 1.3 ($-1$) &                \\
& & & & & & & & & & & & \\
& &                   & \multicolumn{6}{c|}{$M_{\Lambda'\upsilon'\Lambda''\upsilon''}$ ($ea_0$)} & \multicolumn{6}{c}{$M_{\Lambda'\upsilon'\Lambda''\upsilon''}$ ($ea_0$)}  \\[3pt]
& & $\upsilon''$\textbackslash$\upsilon'$
                      & $0$       & $1$       & $2$       & $3$       & $4$       &                & $0$       & $1$       & $2$       & $3$       & $4$       &                \\[3pt]
\multirow{5}{*}{\begin{sideways}$\Lambda' \rightarrow \Lambda''$ \end{sideways}} & \multirow{5}{*}{\begin{sideways}$0 \rightarrow 0$ \end{sideways}}
  & $0$               & 7.2 ($-1$) & 3.6 ($-2$) & 4.5 ($-3$) & 5.6 ($-4$) & 8.2 ($-5$) &                & 7.7 ($-1$) & 6.2 ($-2$) & 3.6 ($-3$) & 2.4 ($-4$) & 2.0 ($-5$) &                \\
& & $1$               &            & 7.3 ($-1$) & 4.7 ($-2$) & 7.5 ($-3$) & 1.1 ($-3$) &                &            & 7.6 ($-1$) & 8.8 ($-2$) & 6.3 ($-2$) & 2.0 ($-3$) &                \\
& & $2$               &            &            & 7.5 ($-1$) & 5.3 ($-2$) & 1.0 ($-2$) &                &            &            & 7.4 ($-1$) & 3.9 ($-2$) & 1.2 ($-2$) &                \\
& & $3$               &            &            &            & 7.7 ($-1$) & 5.6 ($-2$) &                &            &            &            & 7.3 ($-1$) & 3.9 ($-2$) &                \\
& & $4$               &            &            &            &            & 7.8 ($-1$) &                &            &            &            &            & 7.1 ($-1$) &                \\
& & & & & & & & & & & & \\
\hline
& & & & & & & & & & & & \\
& &                   & \multicolumn{6}{c|}{$\sum\,[C]\,k^C_{J'J''}$ (cm$^3$ s$^{-1}$)}            & \multicolumn{6}{c}{$\sum\,[C]\,k^C_{J'J''}$ (cm$^3$ s$^{-1}$)}             \\[3pt]
& & $J''$\textbackslash$J'$
                      & $0$       & $1$       & $2$       & $3$       & $4$       & $5$            & $0$       & $1$       & $2$       & $3$       & $4$       & $5$            \\[3pt]
& & $0$               &           & 7.0 (-11) & 3.3 (-12) & 7.9 (-13) & 1.8 (-13) & 9.7 (-14)      &           & 1.7 (-11) & 1.9 (-11) & 3.1 (-12) & 2.9 (-12) & 7.5 (-13)      \\
& & $1$               & 1.2 (-10) &           & 4.9 (-11) & 1.4 (-12) & 2.8 (-12) & 9.6 (-14)      & 4.9 (-11) &           & 2.5 (-11) & 2.9 (-11) & 5.3 (-12) & 5.4 (-12)      \\
& & $2$               & 2.7 (-12) & 2.5 (-11) &           & 2.4 (-11) & 8.7 (-13) & 2.9 (-12)      & 8.6 (-11) & 4.0 (-11) &           & 2.6 (-11) & 3.2 (-11) & 6.5 (-12)      \\
& & $3$               & 1.5 (-13) & 1.6 (-13) & 5.4 (-12) &           & 1.3 (-11) & 2.1 (-11)      & 1.9 (-11) & 5.9 (-11) & 3.4 (-11) &           & 2.8 (-11) & 3.2 (-11)      \\
& & $4$               & 3.8 (-15) & 3.7 (-14) & 2.3 (-14) & 1.5 (-12) &           & 8.6 (-12)      & 2.1 (-11) & 1.3 (-11) & 4.8 (-11) & 3.3 (-11) &           & 2.6 (-11)      \\
& & $5$               & 1.3 (-16) & 7.6 (-17) & 4.6 (-15) & 1.4 (-13) & 5.1 (-13) &                & 5.8 (-12) & 1.4 (-11) & 1.1 (-11) & 4.0 (-11) & 2.8 (-11) &                \\
& & & & & & & & & & & & \\
\hline
& & & & & & & & & & & & \\
& &                   & \multicolumn{3}{c|}{$\sum\,[D]\,k^D$ (cm$^3$ s$^{-1}$)} & \multicolumn{3}{c|}{$\int \frac{4\pi}{h\nu} \sigma^\gamma(\nu) I_\nu d\nu$ (s$^{-1}$)} & \multicolumn{3}{c|}{$\sum\,[D]\,k^D$ (cm$^3$ s$^{-1}$)} & \multicolumn{3}{c}{$\int \frac{4\pi}{h\nu} \sigma^\gamma(\nu) I_\nu d\nu$ (s$^{-1}$)} \\[3pt]
& &                   & \multicolumn{3}{c|}{8.42 ($-13$)}                       & \multicolumn{3}{c|}{1.17 ($-10$)}                                                      & \multicolumn{3}{c|}{1.16 ($-14$)}                       & \multicolumn{3}{c}{9.80 ($-10$)}                                                       \\
& & & & & & & & & & & & \\
\hline
& & & & & & & & & & & & \\
& &                   & \multicolumn{6}{c|}{$n_J/n({\rm HF})$}                                     & \multicolumn{6}{c}{$n_J/n({\rm CS})$}                                      \\[3pt]
& & $J$               & $0$       & $1$       & $2$       & $3$       & $4$       & $5$            & $0$       & $1$       & $2$       & $3$       & $4$       & $5$            \\[3pt]
& & \ding{182}        & 1.0 (+00) & 2.0 (-04) & 1.2 (-07) & 1.7 (-09) & 1.8 (-11) & 2.9 (-13)      & 2.5 (-01) & 4.8 (-01) & 2.2 (-01) & 3.9 (-02) & 5.3 (-03) & 1.2 (-03)      \\[3pt]
& & \ding{183}        & 1.0 (+00) & 2.0 (-04) & 1.3 (-07) & 2.3 (-09) & 4.3 (-11) & 1.0 (-12)      & 2.5 (-01) & 4.8 (-01) & 2.2 (-01) & 3.9 (-02) & 5.3 (-03) & 1.2 (-03)      \\[3pt]
\multirow{3}{*}{\begin{sideways}$\chi_{\rm nir}$   \end{sideways}} 
& $10^3$ & \ding{184} & 1.0 (+00) & 2.0 (-04) & 1.5 (-07) & 2.4 (-09) & 4.3 (-11) & 1.0 (-12)      & 2.5 (-01) & 4.8 (-01) & 2.2 (-01) & 3.9 (-02) & 5.3 (-03) & 1.2 (-03)      \\
& $10^5$ & \ding{184} & 1.0 (+00) & 2.8 (-04) & 2.1 (-06) & 1.1 (-08) & 5.9 (-11) & 1.0 (-12)      & 2.5 (-01) & 4.8 (-01) & 2.2 (-01) & 4.1 (-02) & 5.4 (-03) & 1.2 (-03)      \\
& $10^7$ & \ding{184} & 9.9 (-01) & 8.2 (-03) & 2.0 (-04) & 1.1 (-06) & 5.3 (-09) & 2.5 (-11)      & 1.3 (-01) & 3.9 (-01) & 3.2 (-01) & 1.2 (-01) & 2.5 (-02) & 4.0 (-03)      \\[3pt]
\multirow{3}{*}{\begin{sideways}$\chi_{\rm opt/UV}$\end{sideways}} 
& $10^2$ & \ding{184} &           &           &           &           &           &                & 2.5 (-01) & 4.8 (-01) & 2.2 (-01) & 3.9 (-02) & 5.3 (-03) & 1.2 (-03)      \\
& $10^3$ & \ding{184} &           &           &           &           &           &                & 2.5 (-01) & 4.8 (-01) & 2.2 (-01) & 4.0 (-02) & 5.3 (-03) & 1.2 (-03)      \\
& $10^4$ & \ding{184} &           &           &           &           &           &                & 2.4 (-01) & 4.7 (-01) & 2.3 (-01) & 4.3 (-02) & 5.8 (-03) & 1.2 (-03)      \\[3pt]
\multirow{3}{*}{\begin{sideways}$\chi_{\rm opt/UV}$\end{sideways}} 
& $10^2$ & \ding{185} & 1.0 (+00) & 2.0 (-04) & 1.4 (-07) & 3.0 (-09) & 7.6 (-11) & 2.0 (-12)      & 2.5 (-01) & 4.7 (-01) & 2.2 (-01) & 4.2 (-02) & 6.0 (-03) & 1.5 (-03)      \\
& $10^3$ & \ding{185} & 1.0 (+00) & 2.1 (-04) & 2.6 (-07) & 9.9 (-09) & 3.8 (-10) & 1.1 (-11)      & 2.1 (-01) & 4.5 (-01) & 2.5 (-01) & 6.2 (-02) & 1.3 (-02) & 4.2 (-03)      \\
& $10^4$ & \ding{185} & 1.0 (+00) & 3.4 (-04) & 1.5 (-06) & 7.8 (-08) & 3.4 (-09) & 9.9 (-11)      & 6.8 (-02) & 2.9 (-01) & 3.2 (-01) & 1.8 (-01) & 7.2 (-02) & 3.0 (-02)      \\
& & & & & & & & & & & & \\
\hline
\end{tabular}
\begin{list}{}{}
\item[$^a$]data from \citet{Webb1968,Johns1959,Meredith1973,Reese2005,Guillon2012}; $^b$data from 
\citet{Lovas1974,Winnewisser1968,Botschwina1985,Turner1992,Lique2007}
\end{list}
\label{Tab-Extension2}
\end{center}
\end{table*}

\begin{table*}[!ht]
\begin{center}
\caption{Same as Table \ref{Tab-Extension1} for HCl and CO. In contrast to other molecules, 
the $\Lambda=1$ state of HCl corresponds to its $C^1\Pi$ (and not $A^1\Pi$) electronic 
configuration.}
\begin{tabular}{c @{\hspace{0.2cm}} c @{\hspace{-0.0cm}} c | c @{\hspace{0.2cm}} c @{\hspace{0.2cm}} c @{\hspace{0.2cm}} c @{\hspace{0.2cm}} c @{\hspace{0.2cm}} c | c @{\hspace{0.2cm}} c @{\hspace{0.2cm}} c @{\hspace{0.2cm}} c @{\hspace{0.2cm}} c @{\hspace{0.2cm}} c @{\hspace{0.2cm}}}
\hline
& &                   & \multicolumn{6}{c|}{$^a$HCl}                                                                                                & \multicolumn{6}{c}{$^b$CO}                                                                                                 \\
\hline
& & & & & & & & & & & & \\
& & $\Lambda$         & \multicolumn{2}{c}{$T_e$ (cm$^{-1}$)} & \multicolumn{2}{c}{$\omega_e$ (cm$^{-1}$)} & \multicolumn{2}{c|}{$B_e$ (cm$^{-1}$)} & \multicolumn{2}{c}{$T_e$ (cm$^{-1}$)} & \multicolumn{2}{c}{$\omega_e$ (cm$^{-1}$)} & \multicolumn{2}{c}{$B_e$ (cm$^{-1}$)} \\[3pt]
& & $0$               & \multicolumn{2}{c}{}                  & \multicolumn{2}{c}{2.990 ($+03$)}          & \multicolumn{2}{c|}{1.059 ($+01$)}     & \multicolumn{2}{c}{}                  & \multicolumn{2}{c}{2.170 ($+03$)}          & \multicolumn{2}{c}{1.931 ($+00$)}     \\
& & $1$               & \multicolumn{2}{c}{7.758 ($+04$)}     & \multicolumn{2}{c}{2.684 ($+03$)}          & \multicolumn{2}{c|}{9.330 ($+00$)}     & \multicolumn{2}{c}{6.508 ($+04$)}     & \multicolumn{2}{c}{1.518 ($+03$)}          & \multicolumn{2}{c}{1.612 ($+00$)}     \\
& & & & & & & & & & & & \\
\hline
& & & & & & & & & & & & \\
& &                   & \multicolumn{6}{c|}{$M_{\Lambda'\upsilon'\Lambda''\upsilon''}$ ($ea_0$)} & \multicolumn{6}{c}{$M_{\Lambda'\upsilon'\Lambda''\upsilon''}$ ($ea_0$)}  \\[3pt]
& & $\upsilon''$\textbackslash$\upsilon'$
                      & $0$       & $1$       & $2$       & $3$       & $4$       &                & $0$       & $1$       & $2$       & $3$       & $4$       &                \\[3pt]
\multirow{5}{*}{\begin{sideways}$\Lambda' \rightarrow \Lambda''$ \end{sideways}} & \multirow{5}{*}{\begin{sideways}$1 \rightarrow 0$ \end{sideways}}
  & $0$               & 8.6 ($-1$) &           &           &           &           &                & 2.9 ($-1$) & 4.0 ($-1$) & 4.2 ($-1$) & 3.8 ($-1$) & 3.0 ($-1$) &                \\
& & $1$               & 5.8 ($-1$) &           &           &           &           &                & 4.2 ($-1$) & 3.3 ($-1$) & 9.1 ($-2$) & 1.2 ($-1$) & 2.3 ($-1$) &                \\
& & $2$               & 4.2 ($-1$) &           &           &           &           &                & 4.2 ($-1$) & 4.2 ($-2$) & 2.5 ($-1$) & 2.9 ($-1$) & 1.7 ($-1$) &                \\
& & $3$               & 1.9 ($-1$) &           &           &           &           &                & 3.3 ($-1$) & 2.2 ($-1$) & 2.7 ($-1$) & 2.5 ($-2$) & 1.8 ($-1$) &                \\
& & $4$               & 1.3 ($-1$) &           &           &           &           &                & 2.2 ($-1$) & 3.3 ($-1$) & 4.9 ($-2$) & 2.4 ($-1$) & 2.2 ($-1$) &                \\
& & & & & & & & & & & & \\
& &                   & \multicolumn{6}{c|}{$M_{\Lambda'\upsilon'\Lambda''\upsilon''}$ ($ea_0$)} & \multicolumn{6}{c}{$M_{\Lambda'\upsilon'\Lambda''\upsilon''}$ ($ea_0$)}  \\[3pt]
& & $\upsilon''$\textbackslash$\upsilon'$
                      & $0$       & $1$       & $2$       & $3$       & $4$       &                & $0$       & $1$       & $2$       & $3$       & $4$       &                \\[3pt]
\multirow{5}{*}{\begin{sideways}$\Lambda' \rightarrow \Lambda''$ \end{sideways}} & \multirow{5}{*}{\begin{sideways}$0 \rightarrow 0$ \end{sideways}}
  & $0$               & 4.3 ($-1$) & 2.8 ($-2$) & 3.1 ($-3$) & 2.2 ($-4$) & 1.2 ($-5$) &                & 4.3 ($-2$) & 4.1 ($-2$) & 2.6 ($-3$) & 1.6 ($-4$) & 8.1 ($-6$) &                \\
& & $1$               &            & 4.9 ($-1$) & 2.0 ($-2$) & 2.0 ($-3$) & 2.0 ($-4$) &                &            & 3.3 ($-2$) & 5.7 ($-2$) & 4.4 ($-3$) & 3.4 ($-4$) &                \\
& & $2$               &            &            & 6.0 ($-3$) & 2.0 ($-2$) & 2.0 ($-3$) &                &            &            & 2.4 ($-2$) & 4.0 ($-2$) & 3.6 ($-3$) &                \\
& & $3$               &            &            &            & 6.1 ($-1$) & 2.0 ($-2$) &                &            &            &            & 1.4 ($-2$) & 4.5 ($-2$) &                \\
& & $4$               &            &            &            &            & 3.4 ($-1$) &                &            &            &            &            & 3.8 ($-3$) &                \\
& & & & & & & & & & & & \\
\hline
& & & & & & & & & & & & \\
& &                   & \multicolumn{6}{c|}{$\sum\,[C]\,k^C_{J'J''}$ (cm$^3$ s$^{-1}$)}            & \multicolumn{6}{c}{$\sum\,[C]\,k^C_{J'J''}$ (cm$^3$ s$^{-1}$)}             \\[3pt]
& & $J''$\textbackslash$J'$
                      & $0$       & $1$       & $2$       & $3$       & $4$       & $5$            & $0$       & $1$       & $2$       & $3$       & $4$       & $5$            \\[3pt]
& & $0$               &           & 7.9 (-12) & 3.0 (-12) & 2.3 (-12) & 1.3 (-12) & 4.5 (-13)      &           & 2.1 (-11) & 2.8 (-11) & 7.8 (-12) & 3.8 (-12) & 2.3 (-12)      \\
& & $1$               & 1.7 (-11) &           & 2.0 (-11) & 7.0 (-12) & 2.4 (-12) & 1.3 (-12)      & 6.1 (-11) &           & 4.1 (-11) & 4.0 (-11) & 1.5 (-11) & 6.4 (-12)      \\
& & $2$               & 6.1 (-12) & 1.8 (-11) &           & 2.3 (-11) & 7.2 (-12) & 1.5 (-12)      & 1.2 (-10) & 6.1 (-11) &           & 4.6 (-11) & 4.3 (-11) & 1.7 (-11)      \\
& & $3$               & 2.6 (-12) & 3.5 (-12) & 1.3 (-11) &           & 2.6 (-11) & 6.9 (-12)      & 3.9 (-11) & 7.2 (-11) & 5.5 (-11) &           & 4.7 (-11) & 4.5 (-11)      \\
& & $4$               & 5.4 (-13) & 4.6 (-13) & 1.5 (-12) & 9.8 (-12) &           & 2.6 (-11)      & 2.0 (-11) & 2.7 (-11) & 5.3 (-11) & 4.9 (-11) &           & 4.8 (-11)      \\
& & $5$               & 5.1 (-14) & 6.9 (-14) & 8.3 (-14) & 7.0 (-13) & 6.8 (-12) &                & 1.1 (-11) & 1.1 (-11) & 2.0 (-11) & 4.3 (-11) & 4.5 (-11) &                \\
& & & & & & & & & & & & \\
\hline
& & & & & & & & & & & & \\
& &                   & \multicolumn{3}{c|}{$\sum\,[D]\,k^D$ (cm$^3$ s$^{-1}$)} & \multicolumn{3}{c|}{$\int \frac{4\pi}{h\nu} \sigma^\gamma(\nu) I_\nu d\nu$ (s$^{-1}$)} & \multicolumn{3}{c|}{$\sum\,[D]\,k^D$ (cm$^3$ s$^{-1}$)} & \multicolumn{3}{c}{$\int \frac{4\pi}{h\nu} \sigma^\gamma(\nu) I_\nu d\nu$ (s$^{-1}$)} \\[3pt]
& &                   & \multicolumn{3}{c|}{2.29 ($-13$)}                       & \multicolumn{3}{c|}{9.80 ($-10$)}                                                      & \multicolumn{3}{c|}{2.28 ($-17$)}                       & \multicolumn{3}{c}{2.60 ($-10$)}                                                       \\
& & & & & & & & & & & & \\
\hline
& & & & & & & & & & & & \\
& &                   & \multicolumn{6}{c|}{$n_J/n({\rm HCl})$}                                    & \multicolumn{6}{c}{$n_J/n({\rm CO})$}                                      \\[3pt]
& & $J$               & $0$       & $1$       & $2$       & $3$       & $4$       & $5$            & $0$       & $1$       & $2$       & $3$       & $4$       & $5$            \\[3pt]
& & \ding{182}        & 1.0 (+00) & 4.9 (-04) & 8.5 (-06) & 8.1 (-07) & 6.1 (-08) & 2.7 (-09)      & 7.3 (-02) & 2.5 (-01) & 3.3 (-01) & 2.2 (-01) & 8.9 (-02) & 2.9 (-02)      \\[3pt]
& & \ding{183}        & 1.0 (+00) & 4.9 (-04) & 8.6 (-06) & 8.2 (-07) & 6.3 (-08) & 2.9 (-09)      & 7.3 (-02) & 2.5 (-01) & 3.3 (-01) & 2.2 (-01) & 8.9 (-02) & 2.9 (-02)      \\[3pt]
\multirow{3}{*}{\begin{sideways}$\chi_{\rm nir}$   \end{sideways}} 
& $10^3$ & \ding{184} & 1.0 (+00) & 4.9 (-04) & 8.7 (-06) & 8.2 (-07) & 6.3 (-08) & 2.9 (-09)      & 7.3 (-02) & 2.5 (-01) & 3.3 (-01) & 2.2 (-01) & 8.9 (-02) & 2.9 (-02)      \\
& $10^5$ & \ding{184} & 1.0 (+00) & 7.5 (-04) & 2.3 (-05) & 8.9 (-07) & 6.3 (-08) & 2.9 (-09)      & 7.1 (-02) & 2.4 (-01) & 3.3 (-01) & 2.2 (-01) & 9.4 (-02) & 3.1 (-02)      \\
& $10^7$ & \ding{184} & 9.7 (-01) & 2.6 (-02) & 1.4 (-03) & 1.3 (-05) & 2.2 (-07) & 4.3 (-09)      & 3.3 (-02) & 9.8 (-02) & 1.6 (-01) & 2.0 (-01) & 2.0 (-01) & 1.5 (-01)      \\[3pt]
\multirow{3}{*}{\begin{sideways}$\chi_{\rm opt/UV}$\end{sideways}} 
& $10^2$ & \ding{184} & 1.0 (+00) & 5.0 (-04) & 9.0 (-06) & 8.5 (-07) & 6.4 (-08) & 3.0 (-09)      & 7.3 (-02) & 2.5 (-01) & 3.3 (-01) & 2.2 (-01) & 9.0 (-02) & 3.0 (-02)      \\
& $10^3$ & \ding{184} & 1.0 (+00) & 6.1 (-04) & 1.3 (-05) & 1.0 (-06) & 7.7 (-08) & 3.4 (-09)      & 7.0 (-02) & 2.3 (-01) & 3.3 (-01) & 2.3 (-01) & 9.8 (-02) & 3.3 (-02)      \\
& $10^4$ & \ding{184} & 1.0 (+00) & 1.7 (-03) & 4.8 (-05) & 3.1 (-06) & 2.1 (-07) & 8.4 (-09)      & 5.3 (-02) & 1.7 (-01) & 2.6 (-01) & 2.6 (-01) & 1.6 (-01) & 7.0 (-02)      \\[3pt]
\multirow{3}{*}{\begin{sideways}$\chi_{\rm opt/UV}$\end{sideways}} 
& $10^2$ & \ding{185} & 1.0 (+00) & 6.5 (-04) & 1.4 (-05) & 1.5 (-06) & 1.5 (-07) & 1.4 (-08)      & 7.2 (-02) & 2.4 (-01) & 3.3 (-01) & 2.2 (-01) & 9.2 (-02) & 3.1 (-02)      \\
& $10^3$ & \ding{185} & 1.0 (+00) & 2.1 (-03) & 6.3 (-05) & 7.4 (-06) & 9.4 (-07) & 1.1 (-07)      & 6.3 (-02) & 2.1 (-01) & 3.1 (-01) & 2.3 (-01) & 1.1 (-01) & 4.2 (-02)      \\
& $10^4$ & \ding{185} & 9.8 (-01) & 1.6 (-02) & 5.6 (-04) & 6.7 (-05) & 8.9 (-06) & 1.1 (-06)      & 3.7 (-02) & 1.2 (-01) & 2.0 (-01) & 2.4 (-01) & 1.9 (-01) & 1.2 (-01)      \\
& & & & & & & & & & & & \\
\hline
\end{tabular}
\begin{list}{}{}
\item[$^a$]data from \citet{Rank1965,Tilford1970,van-Dishoeck1982a,Li2011,Neufeld1994};$^b$data from 
\citet{Macris1985,Eidelsberg1992,Chantranupong1992,Sundholm1995,Goorvitch1994,Kiriyama2001,Flower2001}
\end{list}
\label{Tab-Extension3}
\end{center}
\end{table*}

\subsection{Analytical considerations of the excitation timescales}\label{Excit-times}

The analysis of the main excitation pathways performed in Appendix \ref{Appen-MainExc} shows 
that the pumping processes of \CHp\ can be easily described by a set of analytical formulae. 
This result is inherent to the $X^1\Sigma^+$ ground-state electronic configuration of \CHp\ 
since the associated transition rules favour only one specific excitation pipeline among all others 
(see Fig. \ref{Fig-Main-excitation}). Consequently, the radiative pumping almost solely depends 
on the rotational, vibrational, and vibronic Einstein spontaneous emission coefficients, 
which are proportional to $\nu^3 M^2_{0000}$, $\nu^3 M^2_{0\upsilon'0\upsilon''}$, and 
$\nu^3 M^2_{1\upsilon'0\upsilon''}$ for the far-infrared, near-infrared, and optical/UV pumpings. 
In this section, we focus only on the near-infrared and optical/UV pumpings.

While polyatomic molecules and other ground-state electronic configurations will be treated 
in forthcoming papers (e.g. NH$_3$, OH($^2\Pi$), and CO$^+$($^2\Sigma^+$) in priority), we 
thus select here several $X^1\Sigma^+$ ground-state diatomic molecular species characterized 
by different vibrational and vibronic Einstein coefficients: two light hydrides HF and HCl, and 
three heavy molecules SiO, CS, and CO. Since we aim at providing complete excitation models of 
these chemical compounds, we took into account their main bound electronic configurations:
the $A^1\Pi$ states of CO, CS, and SiO, the $E^1\Sigma^+$ state of SiO characterized by 
high transition probabilities \citep{Elander1973}, and the $C^1\Pi$ state of HCl. Conversely,
we ignore the $A^1\Pi$ states of HF and HCl, which are known to correspond to repulsive 
configurations \citep{Bender1968,Lee1999}, and the $B^1\Sigma^+$ - $X^1\Sigma^+$ system
of HF, whose Franck-Condon factors are found negligible for the $\upsilon' \leqslant 10$, 
$\upsilon''=0$ bands \citep{Alvarino1983}. Lastly, because the rotational, vibrational, 
and electronic structures of these species are clearly separated in energy, and because we 
aim to draw conclusions from their fundamental spectroscopic properties, we reduced their 
effective Hamiltonian to the three prime molecular constants $T_e(\Lambda)$, $\omega_e(\Lambda)$, 
and $B_e(\Lambda)$. 

Within this approximation, the molecules considered are organized as follows:\\[4pt]
\begin{tikzpicture}[scale=0.58]
\draw (  0.0pt,  0.0pt) -- (290.0pt,0.0pt) ;
\draw (  0.3pt, -2.0pt) -- (  0.3pt,2.0pt) ;
\draw ( 10.0pt, -2.0pt) -- ( 10.0pt,2.0pt) ;
\draw ( 17.9pt, -2.0pt) -- ( 17.9pt,2.0pt) ;
\draw ( 24.6pt, -2.0pt) -- ( 24.6pt,2.0pt) ;
\draw ( 30.4pt, -2.0pt) -- ( 30.4pt,2.0pt) ;
\draw ( 35.5pt, -2.0pt) -- ( 35.5pt,2.0pt) ;
\draw ( 40.1pt, -4.0pt) -- ( 40.1pt,4.0pt) ;
\draw ( 70.2pt, -2.0pt) -- ( 70.2pt,2.0pt) ;
\draw ( 87.8pt, -2.0pt) -- ( 87.8pt,2.0pt) ;
\draw (100.3pt, -2.0pt) -- (100.3pt,2.0pt) ;
\draw (110.0pt, -2.0pt) -- (110.0pt,2.0pt) ;
\draw (117.9pt, -2.0pt) -- (117.9pt,2.0pt) ;
\draw (124.6pt, -2.0pt) -- (124.6pt,2.0pt) ;
\draw (130.4pt, -2.0pt) -- (130.4pt,2.0pt) ;
\draw (135.5pt, -2.0pt) -- (135.5pt,2.0pt) ;
\draw (140.1pt, -4.0pt) -- (140.1pt,4.0pt) ;
\draw (170.2pt, -2.0pt) -- (170.2pt,2.0pt) ;
\draw (187.8pt, -2.0pt) -- (187.8pt,2.0pt) ;
\draw (200.3pt, -2.0pt) -- (200.3pt,2.0pt) ;
\draw (210.0pt, -2.0pt) -- (210.0pt,2.0pt) ;
\draw (217.9pt, -2.0pt) -- (217.9pt,2.0pt) ;
\draw (224.6pt, -2.0pt) -- (224.6pt,2.0pt) ;
\draw (230.4pt, -2.0pt) -- (230.4pt,2.0pt) ;
\draw (235.5pt, -2.0pt) -- (235.5pt,2.0pt) ;
\draw (240.1pt, -4.0pt) -- (240.1pt,4.0pt) ;
\draw (270.2pt, -2.0pt) -- (270.2pt,2.0pt) ;
\draw (287.8pt, -2.0pt) -- (287.8pt,2.0pt) ;
\draw ( 40.1pt,  0.0pt) node[below] {$10^6$} ;
\draw (140.1pt,  0.0pt) node[below] {$10^7$} ;
\draw (240.1pt,  0.0pt) node[below] {$10^8$} ;
\draw ( 36.8pt,  0.0pt) node        {$\bullet$} ;
\draw ( 90.9pt,  0.0pt) node        {$\bullet$} ;
\draw (236.4pt,  0.0pt) node        {$\bullet$} ;
\draw (131.3pt,  0.0pt) node        {$\bullet$} ;
\draw (172.3pt,  0.0pt) node        {$\bullet$} ;
\draw (163.7pt,  0.0pt) node        {$\bullet$} ;
\draw ( 36.8pt,  0.0pt) node[above] {\CHp} ;
\draw ( 90.9pt,  0.0pt) node[above] {SiO}  ;
\draw (236.4pt,  0.0pt) node[above] {HF}   ;
\draw (131.3pt,  0.0pt) node[above] {CS}   ;
\draw (172.3pt,  0.0pt) node[above] {HCl}  ;
\draw (163.7pt,  0.0pt) node[below] {CO}   ;
\draw (290.0pt,  0.0pt) node[right] {$\,\,\kappa_{\rm vib}=\omega^3_e(0)M^2_{0100}$} ;
\end{tikzpicture}
\\[4pt]
\begin{tikzpicture}[scale=0.58]
\draw (  0.0pt,  0.0pt) -- (290.0pt,0.0pt) ;
\draw (  2.1pt, -2.0pt) -- (  2.1pt,2.0pt) ;
\draw ( 10.0pt, -2.0pt) -- ( 10.0pt,2.0pt) ;
\draw ( 16.7pt, -2.0pt) -- ( 16.7pt,2.0pt) ;
\draw ( 22.5pt, -2.0pt) -- ( 22.5pt,2.0pt) ;
\draw ( 27.6pt, -2.0pt) -- ( 27.6pt,2.0pt) ;
\draw ( 32.2pt, -4.0pt) -- ( 32.2pt,4.0pt) ;
\draw ( 62.3pt, -2.0pt) -- ( 62.3pt,2.0pt) ;
\draw ( 79.9pt, -2.0pt) -- ( 79.9pt,2.0pt) ;
\draw ( 92.3pt, -2.0pt) -- ( 92.3pt,2.0pt) ;
\draw (102.1pt, -2.0pt) -- (102.1pt,2.0pt) ;
\draw (110.0pt, -2.0pt) -- (110.0pt,2.0pt) ;
\draw (116.7pt, -2.0pt) -- (116.7pt,2.0pt) ;
\draw (122.5pt, -2.0pt) -- (122.5pt,2.0pt) ;
\draw (127.6pt, -2.0pt) -- (127.6pt,2.0pt) ;
\draw (132.2pt, -4.0pt) -- (132.2pt,4.0pt) ;
\draw (162.3pt, -2.0pt) -- (162.3pt,2.0pt) ;
\draw (179.9pt, -2.0pt) -- (179.9pt,2.0pt) ;
\draw (192.3pt, -2.0pt) -- (192.3pt,2.0pt) ;
\draw (202.1pt, -2.0pt) -- (202.1pt,2.0pt) ;
\draw (210.0pt, -2.0pt) -- (210.0pt,2.0pt) ;
\draw (216.7pt, -2.0pt) -- (216.7pt,2.0pt) ;
\draw (222.5pt, -2.0pt) -- (222.5pt,2.0pt) ;
\draw (227.6pt, -2.0pt) -- (227.6pt,2.0pt) ;
\draw (232.2pt, -4.0pt) -- (232.2pt,4.0pt) ;
\draw (262.3pt, -2.0pt) -- (262.3pt,2.0pt) ;
\draw (279.9pt, -2.0pt) -- (279.9pt,2.0pt) ;
\draw ( 32.2pt,  0.0pt) node[below] {$10^{12}$} ;
\draw (132.2pt,  0.0pt) node[below] {$10^{13}$} ;
\draw (232.2pt,  0.0pt) node[below] {$10^{14}$} ;
\draw ( 15.4pt,  0.0pt) node        {$\bullet$} ;
\draw (281.9pt,  0.0pt) node        {$\bullet$} ;
\draw ( 69.3pt,  0.0pt) node        {$\bullet$} ;
\draw (286.0pt,  0.0pt) node        {$\bullet$} ;
\draw (168.7pt,  0.0pt) node        {$\bullet$} ;
\draw ( 15.4pt,  0.0pt) node[above] {\CHp} ;
\draw (281.9pt,  0.0pt) node[above] {SiO}  ;
\draw ( 69.3pt,  0.0pt) node[above] {CS}   ;
\draw (286.0pt,  0.0pt) node[below] {HCl}  ;
\draw (168.7pt,  0.0pt) node[above] {CO}   ;
\draw (290.0pt,  0.0pt) node[right] {$\,\,\kappa_{\rm ele}=T^3_e(\Lambda)M^2_{\Lambda 000},$} ;
\end{tikzpicture}
\\[4pt]
where $\kappa_{\rm vib}$ and $\kappa_{\rm ele}$ are inversely proportional to the lifetimes 
of the excited vibrational and electronic states. According to these schemes and without 
any other considerations, \CHp\ appears to be the species the least sensitive to both the 
near-infrared and optical/UV pumpings. In contrast, the vibrational transition dipole moment 
of HF and the electronic transition dipole moments of SiO and HCl are so high that HF, and 
SiO and HCl are likely candidates to radiative pumpings by near-infrared and optical/UV 
photons. In the following, we compare this simple analytical study with the results obtained 
with MADEX, and we identify the fundamental excitation mechanisms of HF, HCl, SiO, CS, and 
CO, depending on the spectrum of the radiation field pervading a PDR-type medium of constant 
density $\nH = 10^4$ \cc\ and temperature $\TK=100$ K.

The spectroscopic properties and the nonreactive collisional rates of HF, HCl, SiO, CS, and 
CO were implemented in our modified version of the MADEX code, along with their respective
chemical destruction rates and photodissociation rates, both inferred at the border of the 
PDR from the Meudon PDR code. Except for \CHp, we only considered H, \HH, and He as collision 
partners in the computation of the nonreactive nonelastic collisional rates, thus neglecting 
the excitation via electron impact (see, e.g., \citealt{Thummel1992,Itikawa2005}). In the same 
way as for \CHp, MADEX was then run, assuming PDR-type conditions and solar metallicities, in 
four different configurations: 
\ding{182} considering only the excitation by nonreactive collisions, 
\ding{183} considering all collisional excitation processes (i.e. including the chemical pumping), 
\ding{184} considering all line excitation processes (i.e. including the radiative pumping 
of bound vibronic levels), and 
\ding{185} considering all radiative processes (i.e. including the photodissociation)
with different values of the near-infrared and optical/UV scaling factors $\chi_{\rm nir}$ 
and $\chi_{\rm opt/UV}$. The molecular constants, the transition dipole moments, the 
collisional rates, and the results of all models are given in Tables \ref{Tab-Extension1}, 
\ref{Tab-Extension2}, and \ref{Tab-Extension3}. For comparison, we also display in Table 
\ref{Tab-Extension1} the results of this analysis applied to \CHp.

\subsection{Species sensitive to chemical pumping} \label{Sect-Extension-che}

With the chemical destruction rates considered as typical in PDR-type environments (see Tables 
\ref{Tab-Extension1}-\ref{Tab-Extension3}), we find that \CHp\ is the only species, among those 
considered, with low-$J$ levels sensitive to its chemical formation. This concurs with the idea 
that chemical pumping essentially applies to short-lived molecules, and in particular to molecular 
ions with lifetimes below $10^{-2} (10^4/\nH)$ yr (e.g. CO$^+$, \citealt{Stauber2009}). Indeed, 
the chemical pumping has a marginal impact on HCl, SiO, CS, and CO because the timescales of the 
nonreactive collisional excitation of the $J \leqslant 5$ levels are at least one order of 
magnitude lower than their respective destruction timescales.

We note, however, that this last statement should not be generalized to the high-$J$ levels. As was 
already stressed in Sect. \ref{Summary}, the importance of chemical pumping relative to nonreactive 
collisions increases with $J$. For instance, Table \ref{Tab-Extension2} shows that the chemical 
formation takes over the excitation of the rotational levels of HF for $J \geqslant 3$. This suggests 
that this mechanism could play a dominant role for all species, especially in environments with 
a kinetic temperature close to the equivalent temperatures of the rovibrational transitions. A 
direct application could be the interpretation of the high-$J$ CO lines observed during the past
decade in protoplanetary and planteray nebulae, protostellar shocks, and dense and hot PDRs with
the KAO, SOFIA, ISO, and all the instruments onboard the Herschel space observatory \citep{Justtanont1997,
Bujarrabal2010,Benedettini2012,Habart2010}. If the chemical formation efficiently pumps the high-$J$
lines of CO, the high densities $\nH > 10^6$ of the CRL 618 and NGC 7027 molecular clouds inferred 
from CO emission up to the $J=22-21$ transition by \citet{Justtanont1997} and \citet{Bujarrabal2010} 
would be revised to lower values. The chemical pumping could similarly offer an alternative 
understanding of low-mass protostellar outflows where the high-$J$ lines of CO (up to $J=38-37$) 
observed by \citet{van-Kempen2010} with Herschel/PACS have been proposed to trace the UV irradiation 
of the outflow cavity walls and small-scale C-type shocks propagating along the cavity walls 
\citep{Visser2012}.

\subsection{Species sensitive to radiative pumping by near-infrared photons} \label{Sect-Extension-nir}

The comparison of models \ding{183} and \ding{184} shows that HF and HCl are the two species the 
most sensitive to the intensity of the near-infrared radiation field. In contrast, the near-infrared
fluorescence is found to have almost no influence on the populations of the rotational levels of
\CHp\ and SiO and only a marginal effect on those of CS and CO. When cross-compared, these behaviours 
originate from a combination of factors: (1) the nonreactive collisional rates of HCl and HF
(except for the $1-0$ transition) are $\sim 3$ to 10 times lower than those of \CHp, SiO, CS, 
and CO; and (2) the lifetimes of the vibrational levels of HF, HCl, CO, CS, SiO, and \CHp\ increase 
from HF to \CHp\ over two orders of magnitude. It follows that the organization of the species based 
on their sensitivity to radiative pumping of their vibrational states agrees well with the 
classification we inferred from the sole analysis of the $\kappa_{\rm vib}$ values (see Sect. 
\ref{Excit-times}).

Quantitatively, the populations of the $J=1,2,3$ levels of HF increase by a few percent to more
than a factor of 10 for $\chi_{\rm nir}$ varying between $10^3$ and $10^5$. A similar result is 
found for the $J=1,2,3$ levels of HCl, although it requires near-inrared intensities $\sim 10$ 
times higher. Finally, minimum near-infrared radiation fields of $\sim 10^6$ and $\sim 10^7$ times 
that of the local ISRF are needed to efficiently activate the radiative pumping of CS and CO on one 
side and of SiO on the other side.

Overall, the main tracer of near-infrared fluorescence is the $J=2-1$ transition of HF. Unfortunately,
this line has, so far, only been observed in absorption towards Sagittarius B2, \citep{Neufeld1997}.
In turn, the fact that the $J=1$ level of HF is slightly enhanced from $\chi_{\rm nir}=10^3$ to $10^5$ 
is in accordance with the model of \citet{van-der-Tak2012}, who argued that the electron impact excitation
is sufficient to explain the HF $(1-0)$ emission line observed in the Orion Bar interclump medium
with Herschel/HIFI.

\subsection{Species sensitive to radiative pumping of bound electronic states by 
optical and UV photons } \label{Sect-Extension-opt}

The predictions of MADEX concerning the excitation by optical and UV pumping of bound electronic 
states are also found to agree well with the analysis of the $\kappa_{\rm ele}$ values carried 
out in Sect. \ref{Excit-times}. As predicted, SiO and HCl, and to a lesser extent CO, are sensitive 
to the optical and UV radiation field intensities in the configuration \ding{184}, while \CHp\ and 
CS are not. More precisely, we find that the populations of the $J \geqslant 3$ and $J \geqslant 4$ 
levels of SiO and CO increase by more than a factor of 2 via the pumping of their $A^1\Pi$ electronic 
states for an optical/UV radiation field larger than $10^3$ and $10^4$ (in Draine's unit) respectively. 
For $\chi_{\rm opt/UV}=10^4$, we find that this process even induces a population inversion of the 
$J=0$ and $J=1$ rotational levels of SiO. This remarkable behaviour contrasts with that of \CHp\ and 
CS, whose rotational level populations are influenced by optical and UV pumping only for a radiation 
field $10^5$ times larger than the local ISRF.

The applications of the radiative pumping of bound electronic states of SiO extend to a variety 
of astronomical sources. Observed in the Orion Bar and S 140 PDRs by \citet{Schilke2001}, 
the high-$J$ (up to $J=5-4$) and low-$J$ ($J=2-1$ and $J=3-2$) rotational lines of SiO have 
been thought to trace clumps with densities of $2 \times 10^6$ \cc\ and molecular clouds at 
lower densities (down to a few $10^4$ \cc), respectively. With the predictions of Table 
\ref{Tab-Extension1}, we estimate, on the contrary, that the entire SiO emission originates 
from the low-density component and that the high-$J$ lines are confined to the border of the 
PDR where the radiative pumping is maximal. In addition to reproducing the SiO rotational 
diagram, this last result is in concordance with the extended nature of the SiO $(2-1)$ 
emission -- compared to that of SiO $(5-4)$ -- observed in the Orion Bar PDR.

Similar conclusions may be drawn for the high-velocity molecular outflows of class 0 protostars, 
where the high-$J$ (up to $11-10$) rotational lines of SiO associated to the molecular bullets 
distant from the central star are detected with intensities considerably lower than those 
measured in the bullets close to the driving source \citep{Nisini2002,Nisini2007}. Originally 
attributed to density gradients \citep{Nisini2007}, this behaviour may be explained with the 
dilution of the strong FUV radiation field emitted by the young stellar object along the outflow 
cavity (e.g. \citealt{Bruderer2009}).

Finally, the radiative pumping by optical and UV photons could play a major role on the production 
of the SiO masers observed in the region between the photosphere and the dust-growth zone of evolved 
stars envelopes (e.g. \citealt{Gray1995,Cernicharo1993} and references therein). With the transition 
dipole moments of SiO given in Table \ref{Tab-Extension1}, this process is expected to drive the 
excitation of the $\upsilon \geqslant 2$ levels, in contradiction to \citet{Bujarrabal1994}, who
proposed direct infrared pumping of the $\Delta \upsilon=1$ transitions as a global inversion 
process. In a subsequent paper, we will confront the predictions of our complete excitation model
to the forthcoming survey of SiO maser emission (up to $\upsilon=6$) performed on a sample of 50
O-rich evolved stars (in preparation).

\subsection{Species sensitive to their photodissociation}
Finally, Tables \ref{Tab-Extension1}, \ref{Tab-Extension2}, and \ref{Tab-Extension3} show 
that the photodissociation processes have a substantial effect on the rotational diagrams 
of all species considered for $\chi_{\rm opt/UV}$ varying between $10^2$ and $10^4$. 
In particular, the photodissociation, whose rates are derived for an optical/UV flux scaling 
with the standard ISRF, is found to compete with the bound-bound absorption of optical/UV 
photons by SiO, CO, and \CHp, and even to dominate the bound-bound absorption of optical/UV 
photons by HCl and CS.

The species the most sensitive to this mechanism are HF, HCl, and to a lesser extent CS. For 
instance, we find that the photodissociation drives the populations of the $J \geqslant 4$ 
levels of HF and HCl and of the $J \geqslant 3$ levels of CS, providing that the gas is 
illuminated by a UV radiation field $10^2$ and $10^3$ times that of the local ISRF. Moreover, 
since the photodissociation acts as an additional chemical destruction rate, its impact on 
the level populations is similar to the chemical pumping and thus increases with $J$.

These results suggest that the photodissociation may influence the rotational diagrams
of all the species singled out in Sect. \ref{Sect-Extension-che}, \ref{Sect-Extension-nir}, 
and \ref{Sect-Extension-opt}. We note, however, that the importance of the process relative
to the chemical pumping and the radiative pumping of bound states strongly depends on the 
chemical composition of the gas and the spectrum of the optical/UV field. We therefore 
recommend to use models that compute the coupling between the radiative transfer and the 
chemical evolution of the gas to accurately describe the excitation of molecules detected 
in bright PDRs.

\section{Conclusions and perspectives} \label{Conclusion}

We have performed a theoritical investigation of the main excitation and de-excitation mechanisms of 
the methylidyne cation in several astrophysical environments. To do so, we built a new version
of the MADEX radiative transfer model that not only includes all rotational, rovibrational, 
and rovibronic transitions of \CHp, but also takes into account the excitation induced by reactive 
and nonreactive collisions. The model was used to estimate the optical and near-infrared radiation 
fields required to perturb the steady-state rotational level populations of \CHp, and also to provide 
valuable predictions of the intensities of its far-infrared emission lines. The results of MADEX 
were systematically compared with an extensive analytical study of the pumping processes of \CHp, 
which we finally used to extend our predictions to five other $X^1\Sigma^+$ ground-state diatomic 
molecules, HF, HCl, SiO, CS, and CO.

A remarkable agreement was found, throughout this paper, between the results of MADEX and the 
analytical description, thus giving access to the dependence of the excitation processes on the 
main physical conditions of the gas (\nH, \TK, and $I_\nu$) and on the fundamental spectroscopic 
properties of the molecule considered. The key parameters that govern the strength of the radiative 
pumping of the vibrational and bound $^1\Pi$ electronic states of a given species are $\kappa_{\rm vib}=
\omega^3_e(0)M^2_{0100}$ and  $\kappa_{\rm ele}=T^3_e(1)M^2_{1000}$. It follows that 
the far-infrared and submillimetric lines of HF and HCl (SiO, HCl and CO) strongly depend on the intensity 
of the near-infrared (optical/UV) radiation field in astrophysical environments submitted to a 
photon flux at least $10^3 \times (\nH/10^4)$ and $10^4 \times (\nH/10^4)$ times the 
value observed in the solar neighbourhood. Such environments include bright PDRs
where the photodissociation is expected to compete with the line processes in the rotational
excitation of the molecules. The species the most sensitive to the photodissociation are HF and HCl,
whose rotational diagrams are affected by a UV flux as low as $10^2 \times (\nH/10^4)$ times 
that of the local ISRF. In constrast, the distribution of populations among the rotational 
levels of \CHp\ is weakly sensitive to both the radiative pumping of bound states and the 
photodissociation, and is found to strongly depend on the \CHp\ chemical formation.

As a natural consequence, the intensities of the pure rotational lines of \CHp\ observed towards
the Orion Bar and the NGC 7027 PDRs are reproduced with MADEX assuming molecular clouds with
proton densities of $5 \times 10^4$ and $2 \times 10^5$ \cc, respectively. Our estimate of the 
density of the NGC 7027 circumstellar PDR is one to two orders of magnitude below the value 
inferred from traditional excitation models. In addition, we found that the \CHp\ emission 
originates from the interclump medium of the Orion Bar PDR, which contradicts the spatial
correlation found between the \CHp\ $(3-2)$ emission line and the high-density tracers 
such as the rotational $\Lambda$-doublets of OH and high-$J$ rotational transitions of CO.
Consequently, all our results favour molecular clouds with densities significantly lower 
than previously established.

It is foreseeable that similar investigations will have to be carried out for all molecules 
detected in interstellar and circumstellar molecular clouds. In forthcoming papers we will 
address in priority the following problems.
\begin{enumerate}
\item The radiative pumping of the $A^1\Pi$ and $E^1\Sigma^+$ electronic levels of SiO may have 
a dominant role in the production of SiO masers in the envelopes of evolved stars. We intend to 
perform a statistical analysis of the inversion process based on the comparison of the predictions 
of MADEX with the forthcoming observations of the SiO masers in the high vibrational sates (up to 
$\upsilon=6$) in a sample of 50 O-rich evolved stars covering a wide range of mass loss rates.
\item Because the efficiency of the radiative pumping highly depends on the electronic configuration 
of the considered species, we plan to extend the predictions of MADEX to $^3\Sigma^+$ and $^2\Pi$ 
ground-state diatomic molecules. Our goal is to study two molecules that are believed to trace media 
with high density and temperature, CO$^+$ and OH.
\end{enumerate}

\begin{acknowledgements}

We are most grateful to Franck Le Petit, Edith Falgarone, and John Black for providing 
judicious comments on this manuscript and helping to improve its content. We also thank 
the Spanish MICINN for funding support through grants AYA2009-07304 and CSD2009-00038.

\end{acknowledgements}

\bibliographystyle{aa} 
\bibliography{mybib}

\appendix

\section{Analysis of the dominant de-excitation processes of \CHp} \label{Appen-Critic}

\begin{figure}[!h]
\begin{center}
\includegraphics[width=9.0cm,angle=0]{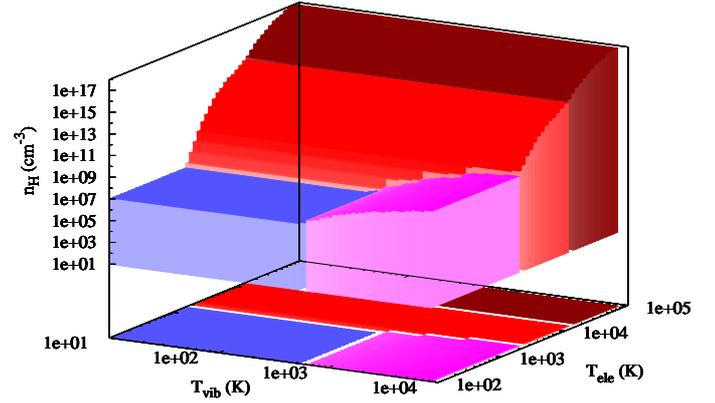}
\caption{Dominant de-excitation process of \CHp\ $J=1$ as a function of the density 
\nH, and the equivalent black body temperatures $T_{\rm vib}$ and $T_{\rm ele}$, 
for $T_{\rm rot} = 2.7$ K. In the blue, purple, red, and brown regions, the de-excitation 
of \CHp\ $J=1$ is dominated by pure rotational, rovibrational, and rovibronic radiative 
transitions and by the photodissociation. In the empty space above, the de-excitation 
occurs through reactive and nonreactive collisions depending on \TK\ and the chemical 
composition of the gas (see main text).}
\label{Fig-crit}
\end{center}
\end{figure}

The concept of critical density -- density at which the nonreactive collisional de-excitation rate 
equals the radiative decay rate -- provides valuable insight into the mechanism that controls the 
population of each level of a molecule depending on the physical conditions of the gas. However, this 
concept is insufficient if state-to-state chemistry is taken into account, or if the wavelengths of 
the radiative de-excitations of the level considered span a wide range of the electromagnetic spectrum, 
as is the case if rovibrational or rovibronic transtitions are considered. We therefore address 
the following question: in this general case, can we derive a set of critical parameters that will 
give us insight into what the steady-sate level populations should be depending on the conditions of 
the gas ?

We consider a species S whose levels are characterized by three quantum numbers only, the 
total orbital angular momentum $\Lambda$, the vibrational momentum $\upsilon$, and the total 
angular momentum $J$. Taking into account all radiative and collisional processes, the density 
of atoms $n_i$ in the level $i$ (with a degeneracy $g_i$) verifies, at steady-state,
{\scriptsize
\begin{equation} \label{eq1}
n_i  = \frac{
\left.\begin{aligned}
\sum\limits_{F} & n_{F_a} n_{F_b} k^F_i + 
\sum\limits_{c} \sum\limits_{j} n_c n_j k^c_{ji} + 
\sum\limits_{j} n_j (A_{ji} + \ngam A_{ji} + \frac{g_i}{g_j} \ngam A_{ij} )
\end{aligned}
\right.
}
{
\left.\begin{aligned}
\sum\limits_{D} & n_D k^D_i +
\sum\limits_{C} \sum\limits_{j} n_C k^C_{ij} + 
\sum\limits_{j} (A_{ij} + \ngam A_{ij} + \frac{g_j}{g_i} \ngam A_{ji} ) +
\int \frac{4\pi}{h\nu} \sigma_i^\gamma(\nu) I_\nu d\nu
\end{aligned}
\right.
}
\end{equation}
and
\begin{equation} \label{eq1b}
n({\rm S}) = \sum\limits_{i} n_i
\end{equation}
}
In these formulae, $n_D$, $n_{F_a}$, and $n_{F_b}$ are the densities of reagents that chemically 
destroy and form the level $i$ at rates $k^D_i$ and $k^F_i$, $n_C$ and $k^C_{ij}$ 
are the abundances of the nonreactive collision partners and the corresponding collisional rates, 
$A_{ij}$ and $\ngam$ are the Einstein spontaneous de-excitation coefficients and the 
directionally averaged photon occupation number\footnote{$\ngam$ is related to the radiative 
energy density $u_\nu$ (in erg cm$^{-3}$ Hz$^{-1}$) and the radiative specific intensity $I_\nu$ 
(in erg cm$^{-2}$ Hz$^{-1}$ sterad$^{-1}$ s$^{-1}$) through : $\ngam = I_\nu c^2/2 h \nu^3$ and 
$u_\nu = 4\pi I_\nu/c$} of the radiation field at the frequency $\nu$ of the $i - j$ transition,
and $\sigma_i^\gamma$ are the photodissociation cross-sections of the level $i$.

To estimate the importance of the different de-excitation processes, we deliberately decomposed 
the radiative term of Eq. \ref{eq1} into three parts that include all transitions: (1) between 
two levels in the same electronic and vibrational state ($\Lambda_i = \Lambda_j$ and $\upsilon_i 
= \upsilon_j$), (2) between two levels in the same electronic state and different vibrational 
states ($\Lambda_i = \Lambda_j$ and $\upsilon_i \ne \upsilon_j$), and (3) between two levels 
in different electronic states ($\Lambda_i \ne \Lambda_j$). We then regrouped all processes 
involving high-energy photons, i.e. the bound-bound rovibronic transitions and the photodissociation. 
In addition, we assumed that the radiation field intensities at the corresponding frequencies 
are described by three black bodies with equivalent temperatures $T_{\rm rot}$, $T_{\rm vib}$, 
and $T_{\rm ele}$ respectively.
Re-arranging Eq. \ref{eq1}, we obtain
{\scriptsize
\begin{equation}\label{eq2}
\left.\begin{aligned}
& n_i = \\[4pt]
& \left(\frac{1}{\displaystyle 1+\frac{n_{{\rm H\,\,cri}, i}}{\nH}} \right)
  \left(\frac{1}{\displaystyle 1+\frac{X_{{\rm cri},i}}{1}} \right)
  \frac{\sum\limits_{F} n_{F_a} n_{F_b} k^F_i}
       {\sum\limits_{D} n_D k^D_i} + \\[4pt]
& \left(\frac{1}{\displaystyle 1+\frac{n_{{\rm H\,\,cri}, i}}{\nH}} \right)
  \left(\frac{1}{\displaystyle 1+\frac{1}{X_{{\rm cri},i}}} \right)
  \frac{\sum\limits_{C} \sum\limits_{j} n_C n_j k^C_{ji}}
       {\sum\limits_{C} \sum\limits_{j} n_C k^C_{ij}} + \\[4pt]
& \left(\frac{1}{\displaystyle 1+\frac{\nH}{n_{{\rm H\,\,cri}, i}}} \right)
  \left(\frac{1}{\displaystyle 1+\frac{Y_{{\rm cri},i}}{1}} \right)
  \frac{\sum\limits_{j\big/\frac{\upsilon_j   =  \upsilon_i}{\Lambda_j = \Lambda_i}} n_j (A_{ji} + \ngam A_{ji} + \frac{g_i}{g_j} \ngam A_{ij} )}
       {\sum\limits_{j\big/\frac{\upsilon_j   =  \upsilon_i}{\Lambda_j = \Lambda_i}} (A_{ij} + \ngam A_{ij} + \frac{g_j}{g_i} \ngam A_{ji} )} + \\[4pt]
& \left(\frac{1}{\displaystyle 1+\frac{\nH}{n_{{\rm H\,\,cri}, i}}} \right)
  \left(\frac{1}{\displaystyle 1+\frac{1}{Y_{{\rm cri},i}}} \right)
  \left(\frac{1}{\displaystyle 1+\frac{Z_{{\rm cri},i}}{1}} \right)
  \frac{\sum\limits_{j\big/\frac{\upsilon_j \neq \upsilon_i}{\Lambda_j = \Lambda_i}} n_j (A_{ji} + \ngam A_{ji} + \frac{g_i}{g_j} \ngam A_{ij} )}
       {\sum\limits_{j\big/\frac{\upsilon_j \neq \upsilon_i}{\Lambda_j = \Lambda_i}} (A_{ij} + \ngam A_{ij} + \frac{g_j}{g_i} \ngam A_{ji} )} + \\[4pt]
& \left(\frac{1}{\displaystyle 1+\frac{\nH}{n_{{\rm H\,\,cri}, i}}} \right)
  \left(\frac{1}{\displaystyle 1+\frac{1}{Y_{{\rm cri},i}}} \right)
  \left(\frac{1}{\displaystyle 1+\frac{1}{Z_{{\rm cri},i}}} \right)
  \left(\frac{1}{\displaystyle 1+\frac{1}{S_{{\rm cri},i}}} \right) \times \\[4pt]
& \frac{\sum\limits_{j\big/\Lambda_j \neq \Lambda_i} n_j (A_{ji} + \ngam A_{ji} + \frac{g_i}{g_j} \ngam A_{ij} )}
       {\sum\limits_{j\big/\Lambda_j \neq \Lambda_i} (A_{ij} + \ngam A_{ij} + \frac{g_j}{g_i} \ngam A_{ji} )},
\end{aligned}
\right.
\end{equation}
}
where each term tends to drag the level populations towards different Boltzmann distributions at 
temperatures \TF, \TK, $T_{\rm rot}$, $T_{\rm vib}$, and $T_{\rm ele}$. In the previous
equation we define five independant parameters, the critical total hydrogen density $n_{{\rm H\,\,cri},i}$
(in \cc), and four dimensionless critical ratios, $X_{{\rm cri},i}$ which depend on the 
abundances relative to H of the reactive and nonreactive collisional partners $[D]$ and $[C]$, and 
$Y_{{\rm cri},i}$, $Z_{{\rm cri},i}$, and $S_{{\rm cri},i}$, which depend on the entire spectrum of 
the radiation field:
\begin{equation} \label{eqncri}
n_{{\rm H\,\,cri}, i} =
\frac{\sum\limits_{j} (A_{ij} + \ngam A_{ij} + \frac{g_j}{g_i} \ngam A_{ji} ) + \int \frac{4\pi}{h\nu} \sigma_i^\gamma(\nu) I_\nu d\nu}
     {\sum\limits_{D} \, [D] \, k^D_{i} + \sum\limits_{C} \sum\limits_{j} \, [C] \, k^C_{ij}}
\end{equation}
\begin{equation} \label{eqXcri}
X_{{\rm cri}, i} =
\frac{\sum\limits_{C} \sum\limits_{j} \, [C] \, k^C_{ij}}
     {\sum\limits_{D} \, [D] \, k^D_{i}}
\end{equation}
\begin{equation} \label{eqYcri}
Y_{{\rm cri},i} = 
\frac{
\left.\begin{aligned}
& {\textstyle \sum\limits_{j\big/\frac{\upsilon_j \neq \upsilon_i}{\Lambda_j = \Lambda_i}} (A_{ij} + \ngam A_{ij} + \frac{g_j}{g_i} \ngam A_{ji} ) \,\, +} \\[-1pt]
& {\textstyle \sum\limits_{j\big/\Lambda_j \neq \Lambda_i} (A_{ij} + \ngam A_{ij} + \frac{g_j}{g_i} \ngam A_{ji} ) + \int \frac{4\pi}{h\nu} \sigma_i^\gamma(\nu) I_\nu d\nu}
\end{aligned}
\right.}
{\sum\limits_{j\big/\frac{\upsilon_j   =  \upsilon_i}{\Lambda_j = \Lambda_i}} (A_{ij} + \ngam A_{ij} + \frac{g_j}{g_i} \ngam A_{ji} )}
\end{equation}
\begin{equation} \label{eqZcri}
Z_{{\rm cri},i} = 
\frac{\sum\limits_{j\big/\Lambda_j \neq \Lambda_i} (A_{ij} + \ngam A_{ij} + \frac{g_j}{g_i} \ngam A_{ji} )+\int \frac{4\pi}{h\nu} \sigma_i^\gamma(\nu) I_\nu d\nu}
{\sum\limits_{j\big/\frac{\upsilon_j \neq \upsilon_i}{\Lambda_j = \Lambda_i}} (A_{ij} + \ngam A_{ij} + \frac{g_j}{g_i} \ngam A_{ji} )}
\end{equation}
\begin{equation} \label{eqScri}
S_{{\rm cri},i} = 
\frac{\sum\limits_{j\big/\Lambda_j \neq \Lambda_i} (A_{ij} + \ngam A_{ij} + \frac{g_j}{g_i} \ngam A_{ji} )}
     {\int \frac{4\pi}{h\nu} \sigma_i^\gamma(\nu) I_\nu d\nu}.
\end{equation}
These quantities prove useful to explore the parameter domain since they indicate the main 
de-excitation mechanism of each level depending on the condition of the gas: through reactive
collisions for $\nH > n_{{\rm H\,\,cri}, i}$ and $X_{{\rm cri},i} < 1$, nonreactive collisions 
for $\nH > n_{{\rm H\,\,cri}, i}$ and $X_{{\rm cri},i} > 1$, rotational radiative transitions 
for $\nH < n_{{\rm H\,\,cri}, i}$ and $Y_{{\rm cri},i} < 1$, vibrational radiative transitions 
for $\nH < n_{{\rm H\,\,cri}, i}$, $Y_{{\rm cri},i} > 1$ and $Z_{{\rm cri},i} < 1$, electronic 
radiative transitions for $\nH < n_{{\rm H\,\,cri}, i}$, $Y_{{\rm cri},i} > 1$, $Z_{{\rm cri},i} > 1$,
and $S_{{\rm cri},i} > 1$, and photodissociation for $\nH < n_{{\rm H\,\,cri}, i}$, 
$Y_{{\rm cri},i} > 1$, $Z_{{\rm cri},i} > 1$, and $S_{{\rm cri},i} < 1$.

When applied with the spectroscopic, collisional and chemical properties of \CHp, the critical 
parameter analysis gives the following results:
\begin{itemize}
\item[$\bullet$] given their short lifetimes, the de-excitation of the vibrational and electronic 
levels of \CHp\ always occurs through spontaneous radiative emission of near-/mid-infrared and 
optical photons.
\item[$\bullet$] The critical ratios $X_{{\rm cri},J} \leqslant 1$ for all rotational levels 
$J \geqslant 1$ as long as $\TK \leqslant 710 \times (2.5-3.0[\HH])^{-1.9}$ K and $\TK \leqslant 
1100 \times ([e^-]/10^{-4})^{-1}$ K. The other critical parameters mainly depend on the density 
of the gas and the intensity of the radiation field.
\end{itemize}
As an example, we therefore display in Fig. \ref{Fig-crit} the limits set by the critical 
parameters analysis for \CHp\ in the level $\Lambda,\upsilon,J=0,0,1$, assuming $[\HH] = 0.5$, 
$[\He] = 0.1$, $[e^-] = 10^{-4}$, $\TK=100$ K, $\TF=\TD=\TK$, and $T_{\rm rot}=2.7$ K. The 
edges at $\nH \sim 10^7$ \cc, $T_{\rm vib} \sim 630$ K, $T_{\rm ele} \sim 1740$ K, and 
$T_{\rm ele} \sim 14400$ K point to the physical conditions for which several de-excitation 
mechanisms have to be taken into account to accurately compute the level populations.

According to Fig. \ref{Fig-crit}, a large UV and optical radiation field seems to be
required for the photodissociation to compete with the bound-bound rovibronic absorption.
However, we emphasize that this result strongly depends on the functional form used to
described the radiation field, i.e. a black body with an effective temperature $T_{\rm ele}$
in the present case. In the standard unshielded interstellar radiation field, the 
photodissociation and the bound-bound rovibronic absorption are found to have comparable
rates and are therefore expected to play similar roles in distributing \CHp\ among
its rotational levels.

\section{Analysis of the dominant excitation processes of \CHp} \label{Appen-MainExc}

\begin{figure*}[!ht]
\begin{center}
\includegraphics[width=19.0cm,angle=0]{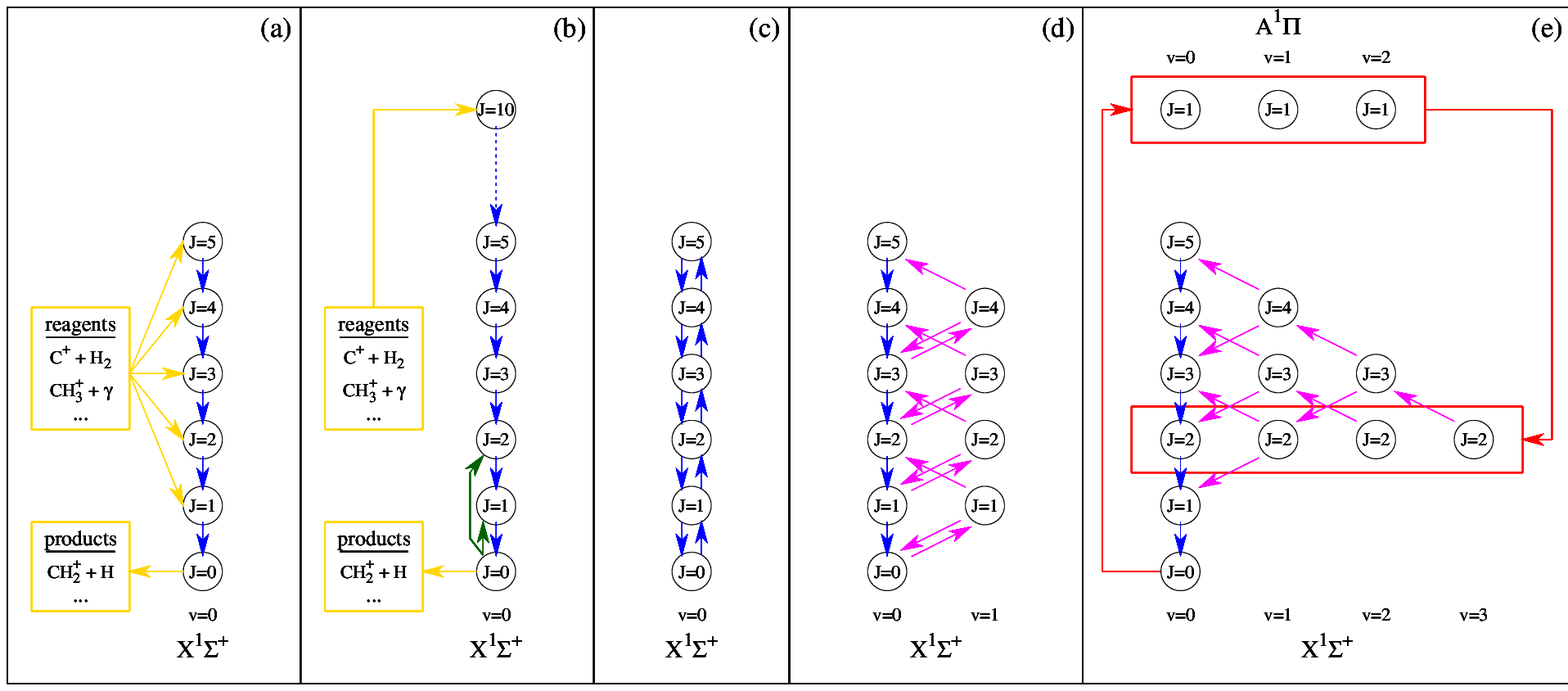}
\caption{Main excitation and de-excitation routes computed with MADEX, assuming $N(\CHp) = 10^{13}$ 
\cq, $\upsilon_{\rm exp} = 1$ \kms, $\nH = 10^4$ \cc, and 
(a) $\TK =  100$ K, $\chi_{\rm fir}=   1$, $\chi_{\rm nir}=      1$, and $\chi_{\rm opt}=   1$, 
(b) $\TK = 1000$ K, $\chi_{\rm fir}=   1$, $\chi_{\rm nir}=      1$, and $\chi_{\rm opt}=   1$,
(c) $\TK =  100$ K, $\chi_{\rm fir}=10^6$, $\chi_{\rm nir}=      1$, and $\chi_{\rm opt}=   1$, 
(d) $\TK =  100$ K, $\chi_{\rm fir}=   1$, $\chi_{\rm nir}=10^{10}$, and $\chi_{\rm opt}=   1$,
(e) $\TK =  100$ K, $\chi_{\rm fir}=   1$, $\chi_{\rm nir}=      1$, and $\chi_{\rm opt}=10^6$.
The photodissociation process was intentionally neglected to focus on the (de)excitation of the 
bound-bound rovibronic transitions. The pure rotational, rovibrational, and rovibronic transitions 
are displayed in blue, purple, and red. Nonreactive and reactive collisional processes are displayed 
in green and yellow. In each case, the levels that do not contribute to the excitation of the pure 
rotational levels are not displayed (e.g. $\Lambda,\upsilon,J = 0,1,0$ in the case (d)). For more 
clarity the rotational emission lines from $J=10$ to $J=5$ are merged into a single dotted arrow 
in case (b), and all vibronic $\Lambda,J = 1,1 \leftarrow 0,0$ absorption lines and $\Lambda,J = 
1,1 \rightarrow 0,2$ emission lines are merged into two single arrows in case (e).}
\label{Fig-Main-excitation}
\end{center}
\end{figure*}

Fig. \ref{Fig-Main-excitation} shows the excitation and de-excitation pathways that drive the
populations of the first five rotational levels of \CHp\ computed with MADEX and neglecting the
photodissociation. Five extreme models were selected: 
(a) $\TK =  100$ K, $\chi_{\rm fir}=   1$, $\chi_{\rm nir}=      1$, and $\chi_{\rm opt}=   1$, 
(b) $\TK = 1000$ K, $\chi_{\rm fir}=   1$, $\chi_{\rm nir}=      1$, and $\chi_{\rm opt}=   1$,
(c) $\TK =  100$ K, $\chi_{\rm fir}=10^6$, $\chi_{\rm nir}=      1$, and $\chi_{\rm opt}=   1$, 
(d) $\TK =  100$ K, $\chi_{\rm fir}=   1$, $\chi_{\rm nir}=10^{10}$, and $\chi_{\rm opt}=   1$,
(e) $\TK =  100$ K, $\chi_{\rm fir}=   1$, $\chi_{\rm nir}=      1$, and $\chi_{\rm opt}=10^6$,
assuming $N(\CHp)=10^{13}$ \cq, $\upsilon_{\rm exp}=1$ \kms, and $\nH = 10^4$ \cc. This figure 
is simplified: for each level, only the processes that together contribute to more than 70 
percent of the total destruction and formation rates are displayed. 

In all cases the de-excitation of the rotational levels occurs via pure rotational transitions, 
as expected from the analytical treatment presented in appendix \ref{Appen-Critic}. Conversely, 
four different excitation regimes are revealed: (a) by reactive collisions for all $J \geqslant 
1$ levels, (b) by chemical pumping of a high-energy level (e.g. $J=10$) followed by the cascade 
through the intermediate $J$ levels, (c) by absorption of far-infrared photons by the $J$-1 
level, (d) by near-infrared pumping of the $\Lambda,\upsilon=0,1$ levels, and (e) by optical 
pumping of the $\Lambda,J=1,1$ levels followed by the cascade through the vibrational levels of 
the $X^1\Sigma^+$ state. Using Eq. \ref{eq1} and taking into account all routes displayed in 
Fig. \ref{Fig-Main-excitation}, we derive at the steady-state equilibrium
\begin{equation}\label{Eq-sse-che}
\left.\begin{aligned}
\frac{n_{00J}}{n(\CHp)} & = \nH \frac{\left( [{\rm H}] \, k^D_{\rm H} + [\HH] \, k^D_{\HH} + 
[e^-] \, k^D_{e^-} \right)}{A_{00J00(J-1)}} \times \\
& \frac{(2J+1) \,\, {\rm exp}(-E_{00J}/\TK)}{\sum\limits_i (2i+1) \,\, {\rm exp}(-E_{i}/\TK)}
\end{aligned} \right.
\end{equation}
for $J \geqslant 1$ in case (a),
\begin{equation}\label{Eq-sse-col}
\left.\begin{aligned}
\frac{n_{00J}}{n(\CHp)} & = \nH \frac{\left( [{\rm H}] \, k^D_{\rm H} + [\HH] \, k^D_{\HH} + 
[e^-] \, k^D_{e^-} \right)}{A_{00J00(J-1)}} \times \\
& \frac{\argmax\limits_{i>J} \left[(2i+1) {\rm exp}(-E_i/\TF)\right]}{\sum\limits_{i} (2i+1) {\rm exp}(-E_i/\TF)}
\end{aligned} \right.
\end{equation}
for $J \geqslant 3$ in case (b),
\begin{equation}\label{Eq-sse-rot}
\left.\begin{aligned}
n_{00J} =  & n_{000} \,\, (2J+1) \,\, \chi_{\rm fir}^{J} \,\, \prod_{i=1}^{J} \langle n_{\gamma,\nu_{ii-1}} \rangle
\end{aligned} \right.
\end{equation}
for $J\geqslant 2$ in case (c),
\begin{equation}\label{Eq-sse-vib-odd}
\left.\begin{aligned}
n_{00J} =  & n_{000} \,\,  \chi_{\rm nir}^{(J+1)/2} \,\, \left( \prod_{i=1}^{(J+1)/2} A_{00(2i-1)00(2i-2)} \right)^{-1} \times \\
& \left( \prod_{i=1}^{(J-1)/2} \frac{\textrm{\scriptsize 4i+1}}{\textrm{\scriptsize 4i-1}}  
\frac{\ngam A_{01(2i)00(2i-1)}}{1+\frac{A_{01(2i)00(2i-1)}}{A_{01(2i)00(2i+1)}}} \right)
\left( 3  \frac{\ngam A_{011000}}{1+\frac{A_{011000}}{A_{011002}}} \right)
\end{aligned} \right.
\end{equation}
for odd $J$-values in case (d),
\begin{equation}\label{Eq-sse-vib-even}
\left.\begin{aligned}
n_{00J} = & n_{000} \,\,  \chi_{\rm nir}^{J/2} \,\, \left( \prod_{i=1}^{J/2} A_{00(2i)00(2i-1)} \right)^{-1} \times \\
& \left( \prod_{i=1}^{(J-2)/2} \frac{\textrm{\scriptsize 4i+3}}{\textrm{\scriptsize 4i+1}} 
\frac{\ngam A_{01(2i+1)00(2i)}}{1+\frac{A_{01(2i)00(2i-1)}}{A_{01(2i)00(2i+1)}}} \right) 
\left( 3 \frac{\ngam A_{011000}}{1+\frac{A_{011000}}{A_{011002}}} \right)
\end{aligned} \right.
\end{equation}
for even $J$-values in case (d),
and 
\begin{equation}\label{Eq-sse-ele}
\left.\begin{aligned}
n_{00J} = & n_{000} \,\,  \frac{\chi_{\rm opt}}{A_{00J00(J-1)}} \times \\ 
& \left( \sum_{\upsilon} 3 \ngam A_{1\upsilon 1000} 
\frac{A_{1\upsilon 10(J-2)2}}{\sum_{\upsilon'} A_{1\upsilon 10\upsilon' 0} }\right) \times \\
& \left( \prod_{i=3}^{J} \frac{1}{1+\frac{A_{0(J-i+1)(i-1)0(J-i)(i-2)}}{A_{0(J-i+1)(i-1)0(J-i)i}}} \right)
\end{aligned} \right.
\end{equation}
for $J\geqslant 2$ in case (e). In all these formulae, $\ngam$ corresponds to the directionally 
averaged photon occupation number of the standard interstellar radiation field at the frequency 
$\nu$ of the radiative transitions.

\end{document}